\documentclass[aps,prx,twocolumn,amsmath,amssymb,superscriptaddress,notitlepage,longbibliography]{revtex4-1}
\usepackage{latexsym}
\usepackage{graphicx}
\usepackage{times,psfrag,subfigure}
\usepackage{enumerate}
\usepackage{amsmath}
\usepackage{dsfont}
\usepackage{dcolumn}
\usepackage{bm,bbm}
\usepackage{hyperref}
\usepackage{color}
\usepackage{latexsym,amsmath,amssymb,bm,euscript, ulem}
\usepackage{dsfont}
\usepackage{textcomp}
\usepackage{tabularx}
\usepackage{setspace}
\usepackage{ctable}
\usepackage{sidecap}
\usepackage{placeins}
\usepackage{threeparttable}
\usepackage{multirow}
\usepackage{pifont}% http://ctan.org/pkg/pifont

\let \oldbm \bm
\renewcommand{\vec}[1]{\oldbm{#1}}

\def\bk{{\vec k}}

\def\bA{{\vec A}}

\def\bs{{\vec s}}
\def\bS{{\vec S}}

\def\bK{{\vec K}}
\def\bq{{\vec q}}

\def\bG{{\vec G}}

\def\bn{{\vec n}}
\def\bm{{\vec m}}

\def\br{{\vec r}}

\newcommand{\inner}[2]{\langle #1|#2\rangle}
\def\tbtau{{\boldsymbol \gamma}}
\def\tbsigma{{\boldsymbol \eta}}
\def\ttau{{\gamma}}
\def\tsigma{{\eta}}
\def \bdelta{{\boldsymbol \delta}}

\def\tr{\mathop{\mathrm{tr}}}
\def\Z{\mathds{Z}}

\def\T{\mathcal{T}}

\def\C{\mathcal{C}}
\def\P{\mathcal{P}}
\def\S{\mathcal{S}}

\def\H{\mathcal{H}}
\def\K{\mathcal{K}}
\def\M{\mathcal{M}}

\def\diag{{\rm diag}}

\def\U{{\rm U}}
\def\u{{\rm U}}
\def\SU{{\rm SU}}

\newcommand{\beq}{\begin{equation}}
\newcommand{\eeq}{\end{equation}}
\newcommand{\beqarray}{\begin{eqnarray}}
\newcommand{\eeqarray}{\end{eqnarray}}

\begin{document}

\title{Ground State and Hidden  Symmetry of Magic Angle Graphene at Even Integer Filling}

\author{Nick Bultinck}
\thanks{N. Bultinck and E. Khalaf contributed equally to this work.}
%\email[]{nickbultinck@berkeley.edu}
\affiliation{Department of Physics, University of California, Berkeley, CA 94720, USA}

\author{Eslam Khalaf}
\thanks{N. Bultinck and E. Khalaf contributed equally to this work.}
%\email[]{eslam15487@gmail.com}
\affiliation{Department of Physics, Harvard University, Cambridge, Massachusetts 02138, USA}

\author{Shang Liu}
%\email[]{sliu@g.harvard.edu}
\affiliation{Department of Physics, Harvard University, Cambridge, Massachusetts 02138, USA}

\author{Shubhayu Chatterjee}
%\email[]{shubhayuchatterjee@berkeley.edu}
\affiliation{Department of Physics, University of California, Berkeley, CA 94720, USA}

\author{Ashvin Vishwanath}
%\email[]{avishwanath@g.harvard.edu}
\affiliation{Department of Physics, Harvard University, Cambridge, Massachusetts 02138, USA}

\author{Michael P. Zaletel}
\email[]{mikezaletel@berkeley.edu}
\affiliation{Department of Physics, University of California, Berkeley, CA 94720, USA}

\date{\today}
\begin{abstract}
    In magic angle twisted bilayer graphene,  electron-electron interactions play a central role resulting in  correlated insulating states at certain integer fillings. Identifying the nature of these insulators is a central question and potentially linked to the relatively high temperature superconductivity  observed in the same  devices. Here we address this question using a combination of analytical strong-coupling arguments and a comprehensive Hartree-Fock numerical calculation which includes the effect of remote bands. The ground state we obtain at charge neutrality  is an unusual ordered state which we call the Kramers intervalley-coherent  (K-IVC) insulator.  In its simplest form, the K-IVC exhibits a pattern of alternating circulating currents  which triples the graphene unit cell leading to an "orbital magnetization  density  wave". Although translation and time reversal symmetry are broken, a combined  `Kramers' time reversal symmetry is preserved. Our analytic arguments are built   on first identifying an  approximate $\U(4) \times \U(4)$  symmetry, resulting from the remarkable properties of the tBG band structure,  which helps select a low energy manifold of states, which are further split to favor the K-IVC. This  low energy manifold is  also found in the Hartree-Fock numerical calculation.  We show that symmetry lowering perturbations can stabilize     other insulators and the semi-metallic state, and discuss the ground state at half filling and a 
    comparison with experiments. 
\end{abstract}
\maketitle

\emph{Introduction}---
In  twisted bilayer graphene (tBG),  two sheets of graphene  twisted by a small angle $\theta$ create a Moir\'e lattice,  resulting in electronic minibands.
For a particular ``magic'' twist angle $\theta \sim 1.05^o$ , theory predicts that the  minibands near charge neutrality (CN) will have minimal dispersion \cite{Santos, MacDonald2011}, and electron-electron interactions play a dominant role. Indeed when the electron filling $\nu$ of these nearly flat bands is varied (completely full/empty bands corresponding to $\nu=\pm4$ electrons per Moir\'e unit cell relative to charge neutrality), insulating states appear at various integer fillings \cite{PabloMott, Dean-Young, efetov}.
The nature of these insulators continue to be debated \cite{Po2018, Thomson18, IsobeFu, KangVafekPRL, Xie2018, CaltechSTM, Liu19, xie2019spectroscopic}.  Furthermore, superconductivity is observed on introducing charge carries into the insulating state \cite{PabloSC, Dean-Young, efetov}. 

Several aspects of the physics of tBG are reminiscent of multi-component quantum Hall systems (e.g. with spin, valley, or layer) where correlated insulators also arise at integer fillings. The driving force there is the  exchange interaction that spontaneously polarizes the electrons into a subset of the components. The Landau-level form of the single particle wavefunctions, which quenches the kinetic energy while preserving their spatial overlap, plays a key role in stabilizing these ferromagnets.
However, the addition of the time reversal symmetry present in tBG, particularly when combined with 180-degree in-plane rotation symmetry  ($C_2$) that effectively enforces time reversal in each valley, opens the door to different orders, including superconductivity, that are absent in the quantum Hall setting. Indeed tBG is one of the few Moir\'e materials that  retains $C_2$ symmetry, which leads to special properties such as unremovable band touchings that double the number of low energy modes. Symmetry-lowering perturbations such as an aligned h-BN  substrate  or weak magnetic fields, are known to induce an integer quantum Hall (IQH) insulator in certain cases \cite{sharpe2019emergent, serlin2019intrinsic}.

In the other canonical model of strong coupling physics,  the Mott-Hubbard  model, symmetry breaking  in the correlated (Mott) insulator is governed by anti-ferromagnetic super-exchange. A pivotal question is whether the single particle subspace defined by tBG leads to insulators that parallel the quantum Hall case, with a cascade of polarized states, or more closely resembles that in  the Hubbard model. We answer this question by considering the structure of Coulomb interactions projected directly into the $\bk$-space continuum model of tBG, including several of the remote bands \cite{Xie2018, Liu19, xie2019spectroscopic}.
While  Mott-Hubbard representation \cite{Thomson18, IsobeFu, KangVafeKPRX, Seo19} are complicated by the topology of the nearly-flat bands \cite{Po2018, StiefelWhitney, Po2018faithful,Song,CaltechSTM, CarrFang19}, one can work directly in the space of the continuum wavefunctions. Here, careful analysis reveals some generic features of the Coulomb matrix elements  which arise from the symmetry and topology of the flat bands.
This analysis allows us to identify both an enlarged  $\U(4) \times \U(4)$ approximate symmetry group and an intervalley-coherent order at neutrality,  missed in previous approaches.

This ``hidden'' symmetry of the model has important phenomenological consequences. Experimentally, many of the basic phenomena, such as the existence of correlated insulators at integer fillings, the location of superconducting domes, and the presence of anomalous  Hall effects, differ from sample to sample. Since the energetics may depend on parameters like the precise twist angle, alignment with the h-BN substrate, and strain, this leads to the sinking feeling that the search for a ``unified'' theory of tBG will become mired in a swamp of microscopic details.
However, in this work we identify a  hierarchy of energy scales in tBG which can naturally unify many of these findings. Due to the remarkable properties of the tBG band structure, we show that the largest energy scales ($15 - 30$ meV) preserve the approximate $\U(4) \times \U(4)$ symmetry which relates a small number of competing symmetry-breaking orders. Smaller effects ($0.2 - 5$ meV) then choose between these orders, and we identify several concrete mechanisms, such as strain or substrate alignment, which can tilt the balance between them.

    \begin{figure}
        \centering
        \includegraphics[width=\columnwidth]{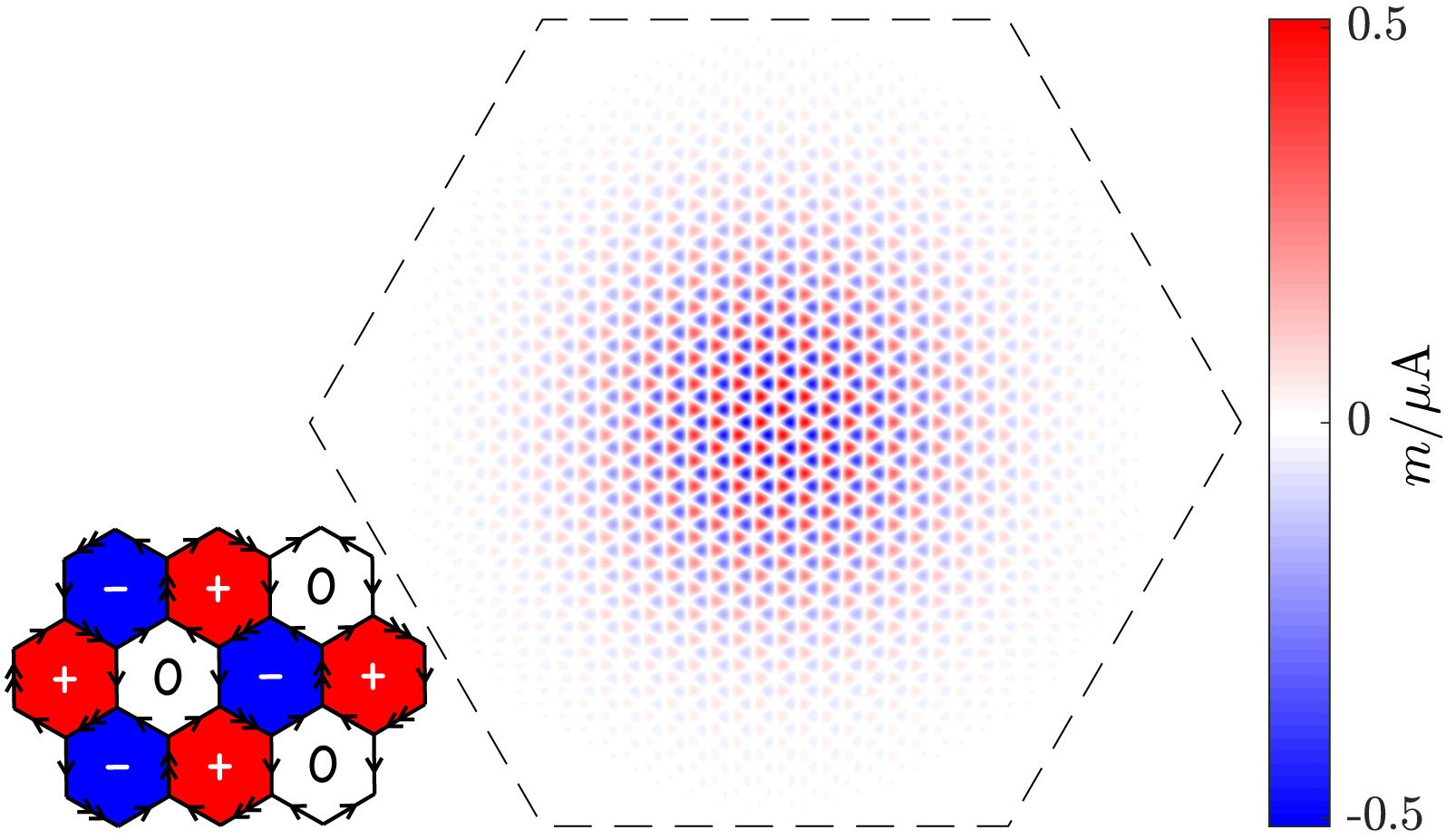}
        \caption{Circulating currents and magnetization of the Kramers intervalley-coherent state (K-IVC)  Similar to a Kekule distortion, spontaneous intervalley-coherence between the $K - K'$ points of the graphene triples the graphene unit cell. The amplitude of the circulating current slowly modulates over the Moir\'e unit cell, shown here as the magnetization density $m(\mathbf{r})$, while preserving the Moir\'e superlattice translations. We show the contribution from a single spin species summed over the two layers; the other spin carries either identical or reversed currents if the K-IVC is a spin singlet or spin `triplet' respectively.
        %, the latter being favored by a ferromagnetic Hund's coupling between valleys.
         Lower-left inset shows an example of the circulating current pattern which retains $C_2\mathcal{T}$ symmetry, at the scale of the \emph{graphene} lattice, in the AA-region of Moir\'e unit cell. } 
        \label{fig:KIVC}
    \end{figure}
    
The primary  focus of this work is to understand the implications of this hierarchy at charge neutrality ($\nu = 0$). 
In certain samples with low twist-angle disorder, an insulating state is  observed in transport at  $\nu = 0$, even in the absence of apparent hBN alignment \cite{efetov}. Scanning tunneling microscopy also finds that the density of states reconstructs at $\nu = 0$, where a gap $\sim$ 15-30 meV opens up \cite{CaltechSTM, RutgersSTM, ColombiaSTM, PrincetonSTM}.
We identify this phase as a new ``Kramers intervalley-coherent'' (K-IVC) state.
In the K-IVC phase (Fig.~\ref{fig:KIVC}), time-reversal is spontaneously broken in each spin component and a pattern of alternating circulating currents develop which triple the graphene unit cell (the Moir\'e unit cell is unchanged). The K-IVC  does not have a net magnetization, but is rather a ``magnetization density wave'' at the wavevector $K$ of graphene's Dirac point. 
Like an anti-ferromagnet, the K-IVC preserves a modified time-reversal symmetry $\mathcal{T}'$  combining the regular (spinless) time reversal $\mathcal{T}$  with a $\pi$ shift in the IVC phase. The new time reversal has the remarkable property that $(\mathcal{T}')^2 = -1$, i.e. it is a Kramers time-reversal symmetry arising from  valley rather than spin.
The presence of $\mathcal{T}'$ leads to Kramers pairing in the spectrum, independent of spin, and may have important implications for the nature of superconductivity  when the K-IVC at $\nu = 0$ is doped. 
Furthermore, restricting to each spin, the K-IVC is a topological insulator, though the protecting $\mathcal{T}'$-symmetry may be strongly broken by the edge (due to broken translation symmetry). 

Before detailing the Hamiltonian, let us briefly summarize the origin of the approximate $\U(4) \times \U(4)$ symmetry. 
The eight flat bands are labeled by spin $s$, valley $\tau$, and a two-fold ``band'' index $\sigma$. Since the bands are quite flat, there is no particular reason that $\sigma$ should label the single-particle eigenbasis.  
Instead, it turns out the two bands can be decomposed into a Chern $C=1$ band and a $C=-1$ band related by $C_2 \mathcal{T}$ symmetry, leading to a total of four $C=1$ and four $C=-1$ bands.
Remarkably, the wavefunctions in the Chern-basis  have a substantial sublattice polarization, i.e. they have a larger projection on one sublattice compared to the other. Thus, we can label them by $\sigma_z = A/B = \pm 1$ with the Chern number $C = \sigma_z \tau_z$. 
Due to this sublattice polarization, the slowly-varying part of the charge density decouples, to a good approximation, into the two Chern components: $n(r) = n_{C=1}(r) + n_{C=-1}(r)$ (otherwise there would be large cross-terms).
The four $C=1$ ($C=-1$) wavefunctions are almost identical up to a permutation of spin and sublattice, so $n(r)$, and hence the interaction, is invariant under separate $\U(4)$ rotations acting on the $C=1 / -1$ components.
The single-particle dispersion and other perturbations then weakly break this symmetry down to the physical one.

This story is in fact highly reminiscent of the QH effect in the zeroth Landau-level (ZLL) of monolayer graphene, which also has a sublattice-valley locking $\sigma_z \tau_z = \textrm{sgn}(B)$ which leads to an approximate $\U(4)$ symmetry.
Indeed, tBG is, in essence,  two time-reversed copies of the ZLL of MLG: $\sigma_z \tau_z = C = \pm 1$, with the tBG flat-band dispersion mapping onto  weak tunneling between the two copies.
This explains why, in the absence of dispersion, and with full sublattice polarization there is then a $\U(4) \times \U(4)$ symmetry coming from each ``ZLL".  Thus, much intuition from the theory of $\U(4)$ quantum-Hall ferromagnetism in MLG \cite{Nomura} can be translated to tBG, albeit with the novel twist of time-reversal symmetry: unlike a single ZLL, unfrustrated Cooper pairs can form from one electron in each copy.

This doubled-ZLL picture also brings us back to the tension between the QH and Hubbard paradigms.
In the end, tBG is a novel hybrid of both: within each copy of the ZLL, the electrons prefer to polarize into a subset of the four components by direct analogy to $\U(4)$ QH ferromagnetism.
However, the tunneling-induced coupling \emph{between} the two ZLLs couples their order-parameters via an anti-ferromagnetic ``$t^2 / U$'' super-exchange. This picks out a submanifold of states comprising of the K-IVC and the valley Hall state. Finally, taking into account the finite sublattice polarization,  the K-IVC which remains a `generalized ferromagnet' is favored relative to the valley Hall state.

\emph{Hamiltonian and symmetries}--- Our starting point is the Bistritzer-Macdonald (BM) \cite{Santos, MacDonald2011} model of twisted bilayer graphene which considers two graphene layers with a relative twist angle $\theta$ coupled via a slowly varying Moir\'e potential.
The interlayer Moir\'e potential is specified by two parameters $w_0$ and $w_1$ denoting intra- and intersublattice coupling, respectively. The ratio  $w_0 / w_1$, which was taken to be 1 in the original BM model, is reduced in realistic samples to about 0.75 due to lattice relaxation effects, which shrink the AA stacking regions relative to the AB regions \cite{Koshino2017,Carr2019}. In the extreme limit where $w_0=0$, an extra chiral symmetry is present which leads to several interesting features including perfectly flat bands at the magic angle \cite{Tarnopolsky}.

Let us now define an extended BM Hamiltonian which includes interactions.
The interaction is taken to be double-gate screened Coulomb interaction with $V_\bq = 2 \pi \tanh(|\bq| d) / \epsilon|\bq|$ where $d$ is the distance to the gate and $\epsilon$ a dielectric constant (similar results are also obtained for the single-gate screened case). Next, we choose a subset of bands of the BM Hamiltonian $h_{\textrm{BM}}$ near charge neutrality labeled by the band index $N_- \leq n \leq N_+$ and assume that all states with $n > N_+$ ($n < N_-$) are empty (full). The projected Hamiltonian has the form
\begin{gather}
\label{Heff}
\H_{\rm eff} = \sum_{\bk \in \rm BZ} c_\bk^\dagger h(\bk) c_\bk - \frac{1}{2A} \sum_\bq V_\bq :\rho_\bq \rho_{-\bq}:, \\
\rho_\bq = \!\!\sum_{\bk \in \rm BZ}\!\! c_\bk^\dagger \Lambda_\bq(\bk) c_{\bk + \bq}, \quad [\Lambda_\bq(\bk)]_{\alpha,\beta} \! = \langle u_{\alpha,\bk} | u_{\beta,\bk + \bq} \rangle
\label{Lambdaqk}
\end{gather}
where $c(\bk)$ is a vector of annihilation operators in the combined index $\alpha, \beta, \dots$ containing spin $s = \uparrow,\downarrow$, valley $\tau = K, K'$ and band $n=N_-,\dots,N_+$ indices, and $u_\alpha(\bk)$ are the eigenstates of the BM Hamiltonian. $A$ is the area and $h(\bk)$ is the single-particle Hamiltonian which includes the BM Hamiltonian as well as band renormalization effects due to the exchange interaction with the filled remote bands (see supplemental material for details \cite{supplementary}) \cite{Liu19, Xie2018,Repellin19}. We neglect electron-phonon interactions as well as the short-distance Coulomb scattering $V_{\bf{K} - \bf{K}'}$ between the Dirac points, both of which are suppressed by powers of the lattice-to-Moir\'e scale $a / L_M \ll 1$. We will refer to these neglected terms as the ``intervalley-Hunds'' terms.

Since the competing $\nu=0$ states are distinguished by their broken symmetries, let us review the symmetries of the extended BM Hamiltonian.
Letting $\sigma_z, \tau_z$ denote sublattice ($A / B$) and valley ($K/K'$), $H_{\textrm{eff}}$ has the following symmetries: (i) $C_2 = \sigma_x \tau_x$ and (ii) $\T = \tau_x \K$ which relate the two valleys, (iii) $C_3 = e^{-\frac{2\pi i}{3} \sigma_z \tau_z}$  which acts within each valley and (iv) $\U(2)_K\times \U(2)_{K'} \simeq \U_C(1) \times \U_V(1) \times \SU(2)_K \times \SU(2)_{K'}$ where $\U_C(1)$, $\U_V(1)$  denote charge conservation, valley charge conservation, and  $\SU(2)_{K,K'}$ represent independent spin rotations in the $K$ and $K'$ valleys. In addition, the BM Hamiltonian has an approximate (v) particle-hole symmetry $\P = i\sigma_x \mu_y \K$ at small angles, where $\mu_i$ are the Pauli matrices acting on the layer index \cite{Hejazi, Song}.

The intervalley Hunds terms, whose magnitude is of the order $J_H\sim 0.2 - 0.5$ meV, break the independent spin rotations in each valley down to the physical global spin rotation symmetry: $\SU(2)_K \times \SU(2)_{K'} \to \SU(2)$. This effect occurs at order $a/L_M\propto \theta$. Furthermore, umklapp processes which scatter three electrons between the two valleys (either due to phonons, or higher-order Coulomb scattering)  break $U_V(1)$ down to $\mathbb{Z}_3$, and are suppressed by a further factor of $\theta^2$ \cite{aleiner2007spontaneous, WuXu19}.

\emph{Hartree-Fock mean-field}--- In the Hartree-Fock (HF) method, we solve for the set of self-consistent ground state Slater determinant states characterized by the one-electron density matrices $P_{\alpha,\beta}(\bk) = \langle c_\alpha^\dagger(\bk) c_\beta(\bk) \rangle$. Similar to Refs. \cite{Xie2018,CaltechSTM,xie2019spectroscopic}, we take both the flat bands and a range of remote bands around charge neutrality into account. However, in contrast to previous studies \cite{Xie2018,CaltechSTM, Liu19,xie2019spectroscopic}, we allow for coherence between the two valleys which spontaneously breaks the $U_V(1)$ symmetry (see also Ref.~\cite{Po2018} for an early suggestion of a different IVC order motivated on phenomenological grounds).  Further details of our procedure are provided in the supplemental material.

    \begin{figure}
        \centering
        \includegraphics[width=0.48\textwidth]{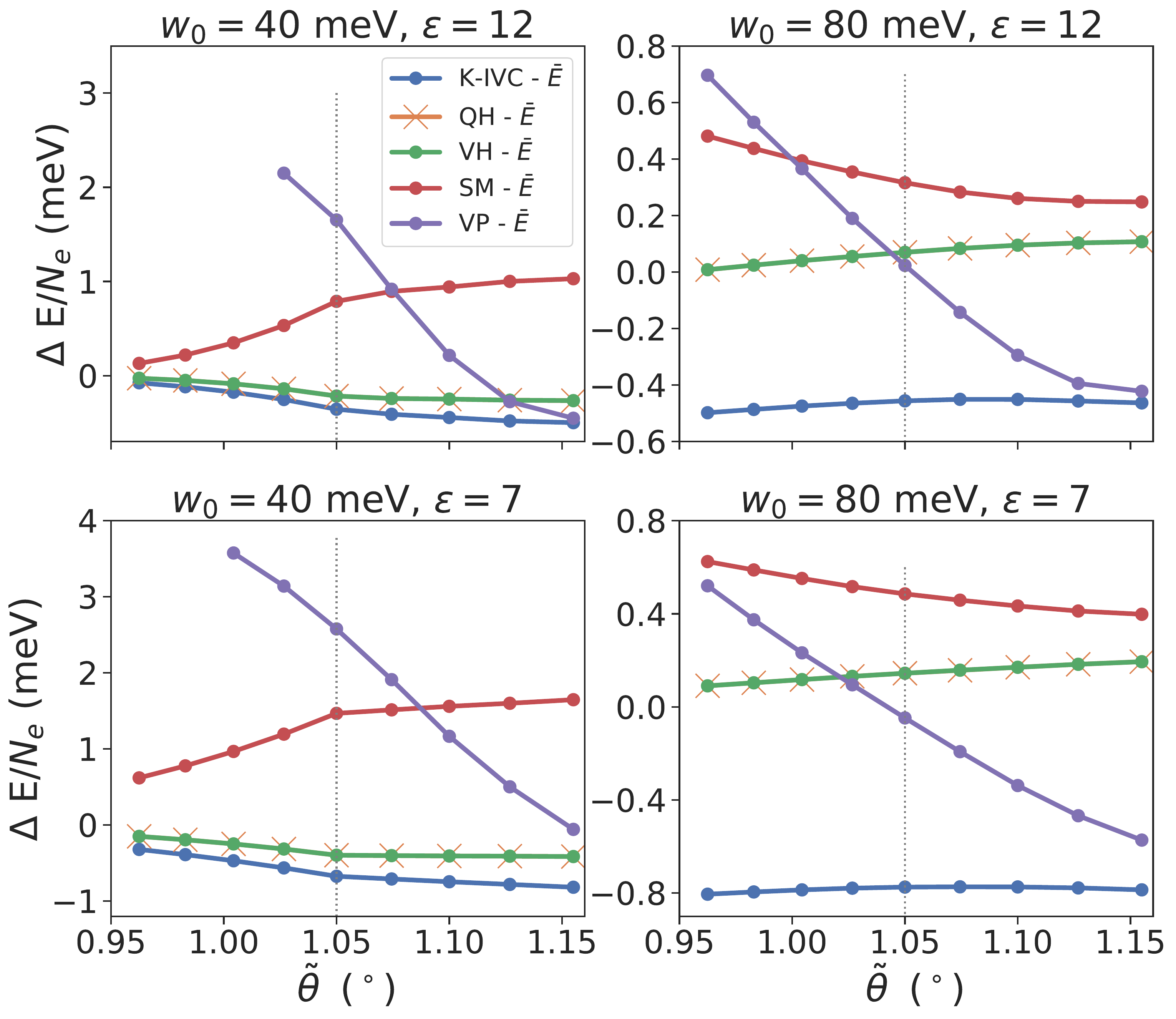}
        \caption{Energies per electron at charge neutrality in the K-IVC, QH, VH, SM and VP states relative to the average energy of the K-IVC, QH, VH and SM (denoted as $\bar{E}$). 
        Results are obtained at $\theta = 1.05^\circ$ as a function of  $100 \leq w_1 \leq 120$ [meV] (x-axis) for $w_0 = 40, 80$ meV and $\epsilon = 7, 12$.  For convenience we \emph{define} $\tilde{\theta} = 1.05^{\circ}\times (110/w_1\,[\text{meV}])$ in order to convert $w_1$ to a qualitatively equivalent angle. 
         The dashed vertical line shows the first magic angle.
         Results were obtained using six Moir\'e bands per spin and valley, and a $24\times 24$ momentum grid. Note that the energies of the VH and QH states are numerically identical.}
        \label{HFnumerics}
    \end{figure}
    
The numerical results at CN ($\nu = 0$) are given in Fig.~\ref{HFnumerics} for fixed $\theta = 1.05^o$, $\epsilon = 7, 12$,  and $w_0 = 40, 80$ meV as a function of $w_1$.
Since the magic angle condition depends on the ratio $w_1/\theta$ \cite{MacDonald2011}, this is approximately equivalent to changing $\theta$.
We exploit this fact to plot the HF energies as a function of an ``effective'' angle $\tilde{\theta}\equiv 1.05^{\circ}\times (110\, /w_1 [\text{meV}])$, where $w_1=110$ meV is the magic angle condition for the parameters we have used.
From comparison with ab-initio methods, the magnitude of the inter-layer tunneling terms are estimated to be $w_1 \sim 110$ meV and $w_0 \sim 80$ meV \cite{MacDonald2011,Koshino2017,Carr2019}.
Here, we consider a range of values of $w_{0/1}$ which can be far from these estimates as this provides valuable information when comparing numerical results with our analytical findings below.

Depending on the initial condition or which symmetries are explicitly enforced, we find several self-consistent solutions which can be grouped into three categories:  (i) a semimetallic (SM) state which preserves $C_2$, $\T$, and $\U_V(1)$ but \emph{may} break $C_3$ 
(this state can be understood as a renormalized version of the BM semi-metallic band structure); (ii) a quantum hall (QH) insulator with Chern number $\pm 4$ which breaks $\T$ but preserves $C_2$ and $\U_V(1)$; and (iii) several insulating states with Chern number 0,  including valley-Hall (VH) state, which breaks $C_2$ but preserves $\T$ and $\U_V(1)$, valley-polarized (VP) state \footnote{Depending on the parameters, the VP state can also be metallic as a result of the interaction between the remote bands and the active bands.}, which breaks $\T$ and $C_2$ but preserves $C_2 \T$ and $\U_V(1)$, and an intervalley coherent (IVC) state which breaks $\T$ and $\U_V(1)$ but preserves the combination $\T' = \tau_y \K$ which acts as a spinless Kramers time-reversal symmetry between valleys. Unlike previously studied IVC states in TBG \cite{Bultinck19} and related Moir\'e materials \cite{Zhang2018, Lee19}, this Kramers IVC (K-IVC) takes place between wavefunctions which have the {\em same} Chern number, thus evading the energy penalty associated with vortices in the order parameter \cite{Bultinck19}.

\begin{figure}[t]
        \centering
        a)
        \includegraphics[width=0.23\textwidth]{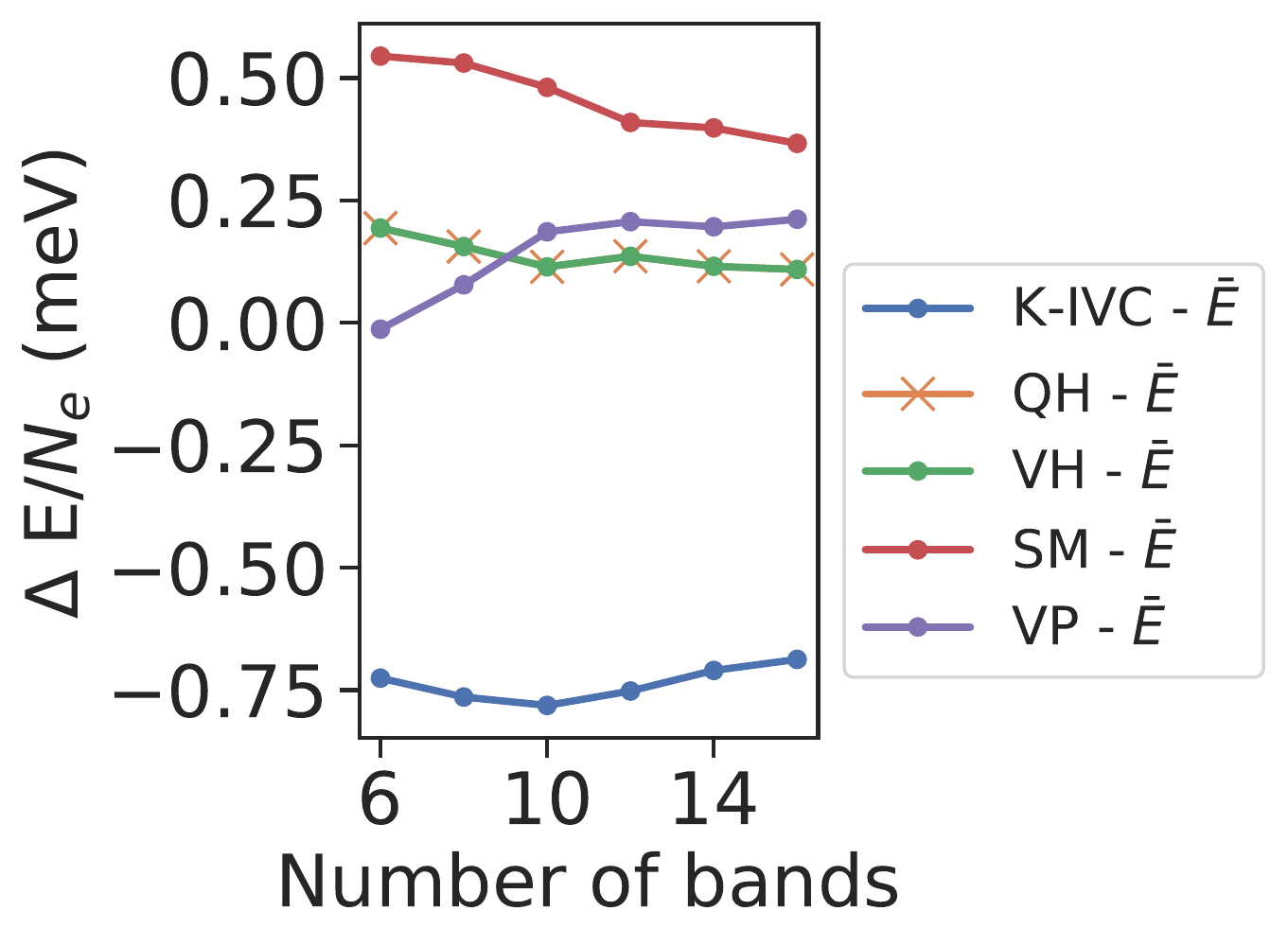}
        b)
        \includegraphics[width=0.2\textwidth]{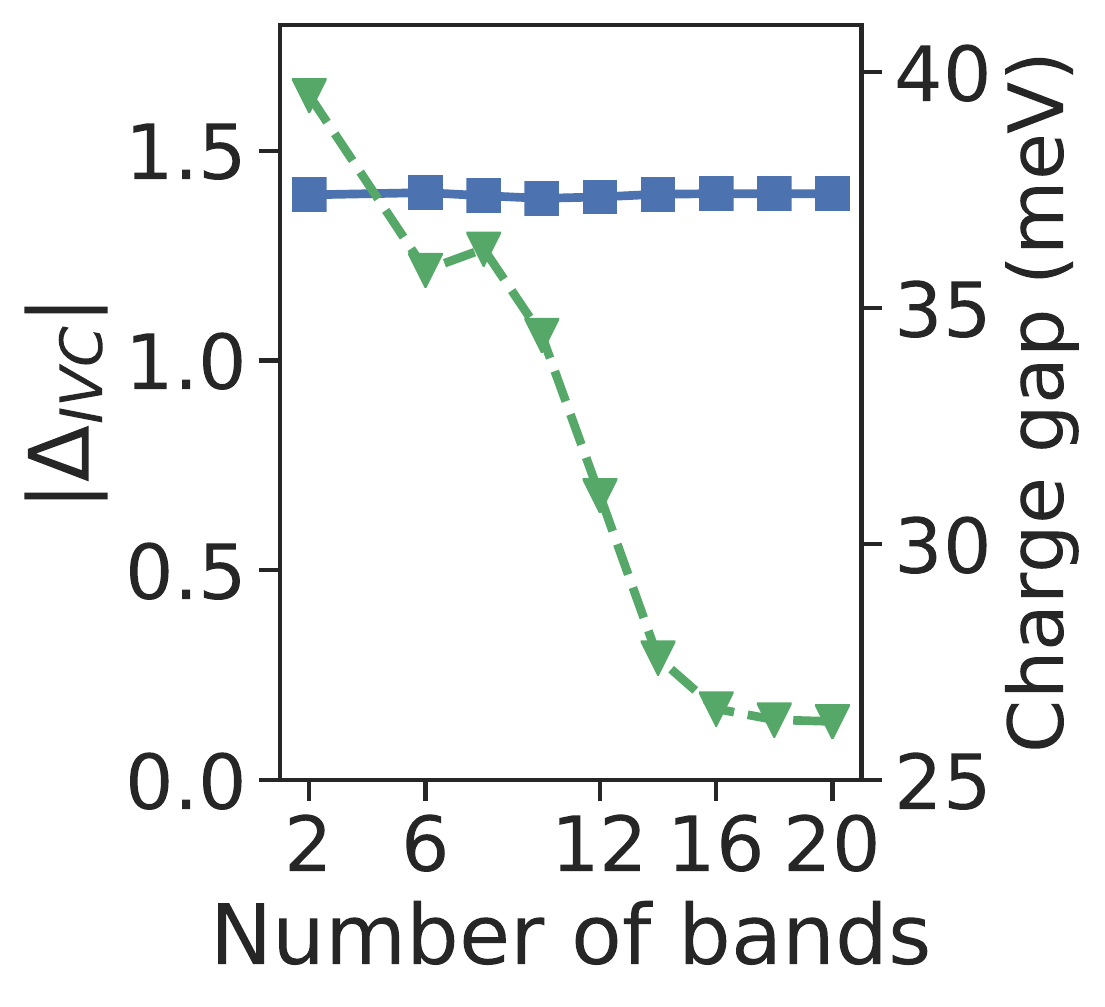}
        \caption{ (a) Energies per electron in the flat bands of the K-IVC, QH, VH, SM and VP states relative to the average energy of the K-IVC, QH, VH and SM (denoted as $\bar{E}$), as a function of the number of bands per spin and valley kept in the Hartree-Fock numerics. (b) IVC order parameter $|\Delta_{\rm IVC}| = \sum_k \tr(P_{\rm IVC}(\bk)^2)^{1/2}/N_M$, where $P_{\rm IVC}(\bk)$ is the $\U_V(1)$-breaking part of $P(\bk)$ and $N_M$ the number of Moir\'e unit cells, (left, blue squares) and the charge gap (right, green triangles) at charge neutrality as a function of the number of bands per spin and valley. The results in both (a) and (b) were obtained on a $12\times 12$ momentum grid with $\theta=1.05^\circ$, $w_0=80$ meV, $w_1=110$ meV and $\epsilon=7$. Note that the energies of the VH and QH states are numerically identical.}
        \label{NrBands}
\end{figure}

The competition between the VH, VP, QH, and SM states, which were all found in previous mean field studies \cite{Xie2018,CaltechSTM, Liu19}, is very sensitive to the values of ($w_0$, $w_1$). This explains why these studies, all of which assumed unbroken $\U_V(1)$ symmetry, did not agree on the nature of the ground state. On the other hand, the $\U_V(1)$-breaking K-IVC state is always the lowest energy state regardless of the values of $w_0$, $w_1$ and $\epsilon$. 
Another salient feature is that the competition  between the K-IVC, QH, and VH is closest when $w_0 \to 0$, but is lifted in favor of the K-IVC for larger $w_0$. The reason will become clear from our analysis of the approximate symmetries. 

The HF numerics shown in Fig.~\ref{HFnumerics} were obtained by keeping six bands per spin and valley, but more generally we find that mixing between the flat and remote bands has only a quantitative effect over the range of parameters considered. 
In particular, the K-IVC remains the ground state as more bands are included, and the magnitude of the IVC order parameter remains almost unchanged (Fig.~\ref{NrBands}), indicating the symmetry-breaking occurs predominantly in the flat bands. The charge gap decreases quantitatively as more bands are included, but saturates at a value of $\sim 26$ meV when sixteen bands per spin and valley are taken into account, and a value of $\epsilon = 7$ is used. As a result, our numerical results can be reproduced to a good degree of accuracy within the two-band projection of Ref.~\cite{Liu19}, where the effect of the remote bands is incorporated only via the exchange-renormalization of $h(\bk)$.

 \begin{figure}[t]
        \centering
        a)
        \includegraphics[width=0.25\textwidth]{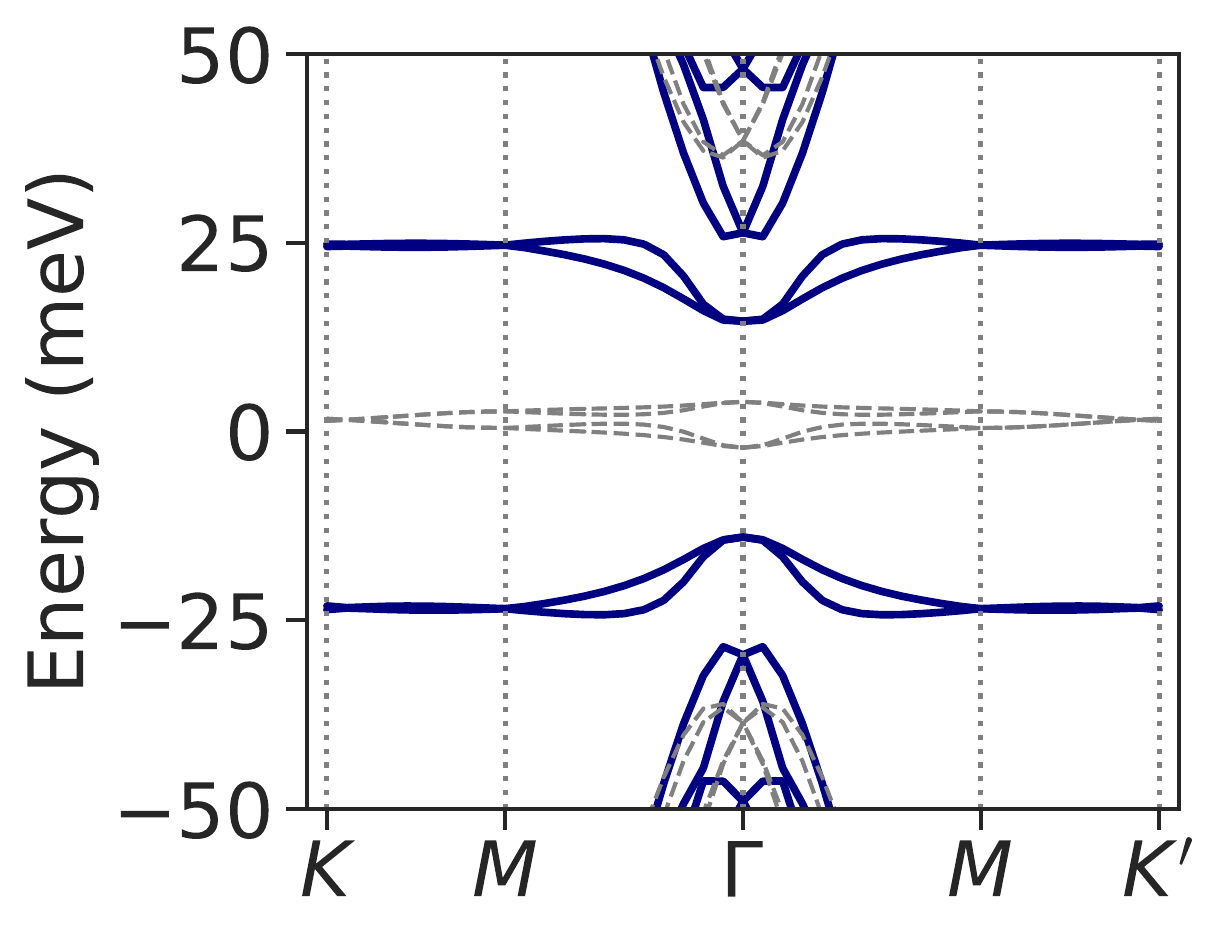}
        \hspace{0.2 cm}
        b)
        \includegraphics[width=0.17\textwidth]{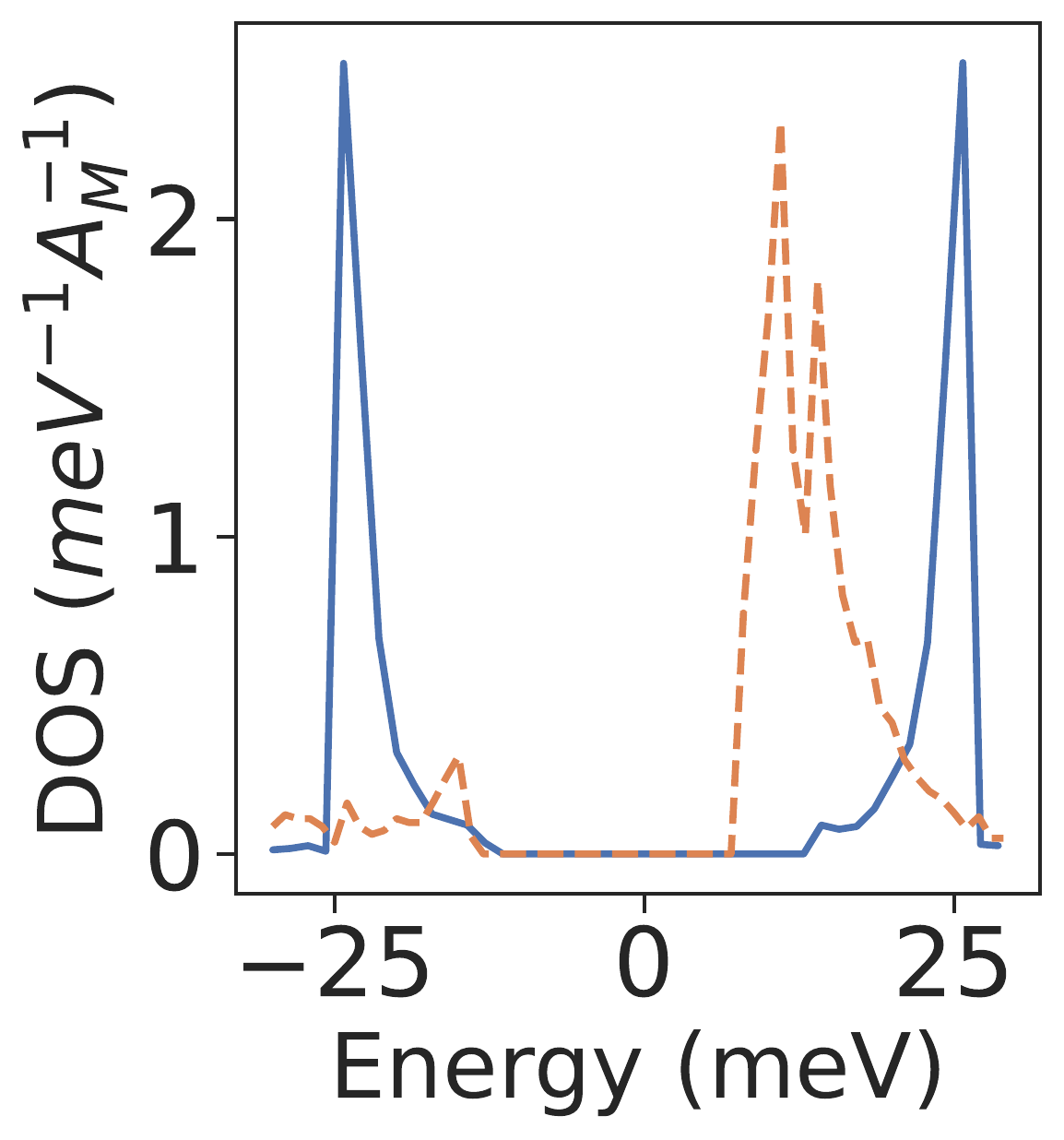}
        \caption{(a) HF band spectrum of the K-IVC state that solves the self-consistency equations when six bands per spin and valley are used. The parameters were $\theta = 1.05^\circ$, $w_0=78$ meV, $w_1=105$ meV and $\epsilon=9.5$. The gray dashed lines correspond to the original BM band spectrum. (b) Density of states (DOS) of the K-IVC state at charge neutrality (full, blue line) and the self-consistent HF solution with empty flat bands, i.e. at filling $\nu=-4$ (dashed, orange line). $A_M$ is the area of the Moir\'e unit cell.
        }
        \label{HFbands}
    \end{figure}
    
A better intuition for the symmetry-breaking phases in Fig.~\ref{HFnumerics} can then be obtained by restricting to the flat bands, where $P(\bk)$ is an 8 by 8 matrix which we parameterize as $P(\bk) = \frac{1}{2}(1 + Q(\bk))$, with $Q(\bk)^2 = 1$ and $\tr Q(\bk) = 2 \nu$.
Furthermore, rather than working in the basis which diagonalizes $h_{\textrm{BM}}$
, it is  convenient to work in the sublattice-polarized basis which diagonalizes the sublattice operator $\sigma_{mn}(\bk) = \langle u_n(\bk)|\sigma_z|u_m(\bk) \rangle$, with $n,\,m\in \{1,\,2\}$ restricted to the two flat bands. This basis is well-defined as long as the eigenvalues of the matrix $\sigma(\bk)$ are non-zero, indicating finite sublattice polarization. In the supplemental material we check that this is indeed the case. The 8 flat bands are then labeled by $s_z = \uparrow / \downarrow, \tau_z = K / K',  \sigma_z = A / B$.
A crucial feature of this basis is that each band carries a quantized Chern number $C = \tau_z \sigma_z$ \cite{Zou2018,Tarnopolsky,XiDai, Bultinck19}.

With this basis in hand, we can concisely summarize the competing insulators: $Q_{\textrm{QH}} = \sigma_z \tau_z = C$ (which explains its net Hall conductance); $Q_{\textrm{VP}} = \tau_z$; $Q_{\textrm{VH}} = \sigma_z = \tau_z C$ (which explains its valley-Hall conductance); and finally 
\begin{equation}
Q_{\textrm{K-IVC}} = \sigma_y \left[ \cos(\theta_{\textrm{IVC}}) \tau_x + \sin(\theta_{\textrm{IVC}}) \tau_y  \right]
\end{equation}
which was found to be the ground state at charge neutrality for the entire parameter range that was studied.
Under (graphene-scale) lattice translations, the K-IVC order parameter transforms as $\theta_{\textrm{IVC}} \to \theta_{\textrm{IVC}} + \frac{2 \pi}{3}$, while under spinless $\mathcal{T}$, $\theta_{\textrm{IVC}} \to \theta_{\textrm{IVC}} + \pi$.
In addition to the spin-singlet variant of the K-IVC state discussed here, there are other K-IVC states with different spin structures which are all degenerate on the level of $\H_{\rm eff}$. These will be discussed below in the sections containing our analytical results.

\emph{Enlarged $\U(4) \times \U(4)$ symmetry}--- Below, we will show how a large $\U(4)\times \U(4)$ symmetry appears in the pure interaction model (i.e. with no dispersion) in the chiral limit. We will begin by showing that even away from the chiral limit, the flat-band-projected interaction term has an enhanced $\U(4)$ symmetry. Next we will then show that the  chiral model also has a different enhanced $\U(4)$ symmetry, even when dispersion is included. Combining these we will obtain a large $\U(4)\times \U(4)$ symmetry for the  chiral model in the absence of dispersion. 

Motivated by the numerical result, we are going to restrict ourselves in the following to the two flat bands (per spin and valley) and rewrite the interacting Hamiltonian (\ref{Heff}) as
\begin{gather}
\H_{\rm eff} = \sum_\bk c_\bk^\dagger \tilde h(\bk) c_\bk +  \frac{1}{2A} \sum_{\bq} V_\bq \delta \rho_\bq \delta \rho_{-\bq} + \text{const.}\\ \delta \rho_\bq = \rho_\bq - \bar \rho_\bq, \qquad \bar \rho_\bq = \frac{1}{2} \sum_{\bG,\bk} \delta_{\bG,\bq} \tr \Lambda_\bG(\bk)
\label{HProj}
\end{gather}
where the interaction term differs from (\ref{Heff}) by an exchange term due to normal ordering as well as the subtraction of the average charge density at neutrality $\sum_\bq \bar \rho_\bq$ (see supplemental material for details). The resulting density operator $\delta{\rho}_{\mathbf{q}}$ is exactly odd under particle-hole, and hence $\tilde{h}$ and the interaction are \emph{separately} particle-hole symmetric ($\sum_\bq \bar \rho_\bq$ is the total charge-density of the flat bands).

Let us first consider the limit where sublattice polarization is not saturated, i.e. chiral symmetry is not present $w_0 \neq 0$. 
Now, the particle-hole symmetry of the projected Hamiltonian (\ref{HProj}) has important consequences. This follows from the observation that a $\P \T$ symmetry (which flips energy but not momentum) is equivalent, within a perfectly flat band (i.e on {\em ignoring} the single particle dispersion), to a single particle {\em unitary} symmetry since it leaves the space of eigenstates invariant. In our model, the gauge can be chosen such that the $\P \T$ symmetry has the following simple form in the flat band projected basis (see supplemental material)
\beq
i\P \T = \tau_y \sigma_y.
\eeq
$\P \T$ acts locally in space and momentum but exchanges valley and sublattice, relating flat-bands with the same Chern number $C = \tau_z \sigma_z$. Thus, if we neglect the dispersion term $\tilde h$, we find that the $\U(2)_K \times \U(2)_{K'}$ of the Hamiltonian is enlarged to a $\U(4)_{\P \T}$ symmetry whose generators are $\{t^a, t^a \sigma_y \tau_y\}$ where $t^a$ are the 8 (sublattice and valley diagonal) generators $t^a = \{s_\mu, \tau_z s_\mu \}$ of $\U(2)_{K} \times \U(2)_{K'}$ and $\mu=\in \{0,\,1,\,2,\,3\}$. This unitary symmetry is broken by the dispersion term $\tilde h$ which anticommutes with the extra generators $t^a \sigma_y \tau_y$.

Another limit where the symmetry of the Hamiltonian is enhanced is the chiral limit $w_0 = 0$ \cite{Guinea,Tarnopolsky}, where the BM Hamiltonian has an extra chiral symmetry $\S = \sigma_z$,  $\{\S, H_{\textrm{BM}} \} = 0$, leading to complete sublattice polarization. In this case, we can combine $\P \T$ symmetry with $\S$ to obtain a $\Z_2$ unitary symmetry $R$ given by
\begin{equation}
R = \P \T \S = \tau_y \sigma_x
\end{equation}
Similar to $\P \T$, $R$ acts locally in space and momentum but exchanges valley and sublattice, relating bands with the same Chern number $C = \tau_z \sigma_z$. Its existence enlarges the symmetry of the model to $\U(4)_R$ whose generators are $\{t^a, t^a R\}$. It is important to notice that this $\U(4)_{R}$ symmetry is different from the $\U(4)_{\P \T}$ symmetry discussed earlier. In addition, the $\U(4)_{R}$ symmetry is preserved on including the dispersion $\tilde h$ and does not rely on the flat band projection, i.e. it is a symmetry of the full Hamiltonian in the chiral limit.

\begin{figure}
\centering
\includegraphics[width = 0.48 \textwidth]{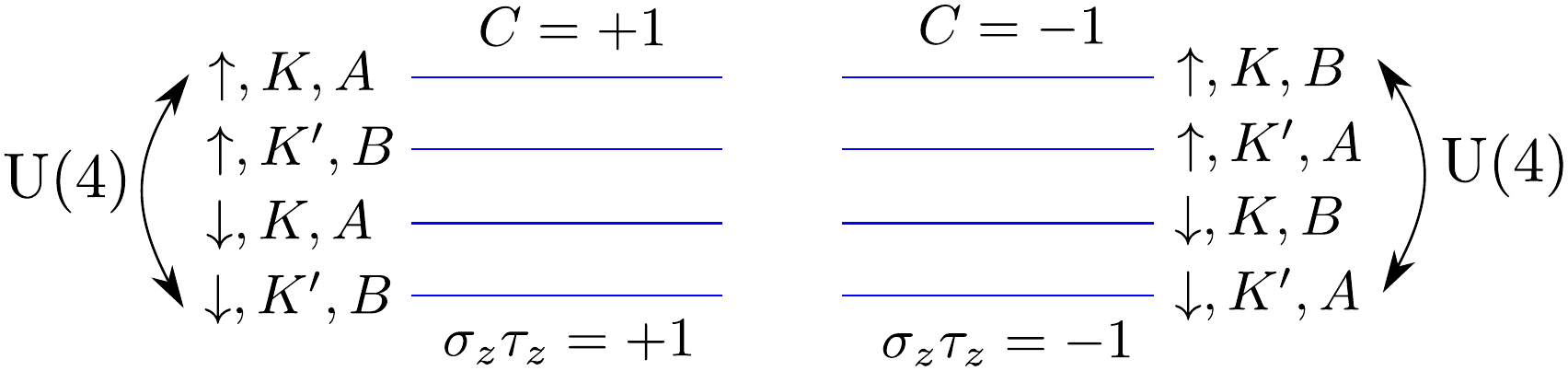}
\\
\vspace{0.3 cm}
%{\bf Hierarchy of energy scales}
%\small
\begin{tabular}{c|c|c}
\hline \hline
Term & Symmetry & Energy scale \\ \hline
$U_S$ & $\U(4) \times \U(4)$ & 15-25 meV\\
$t_S$ & $\U(4)_{R}$ & 4-6 meV \\
$U_A$ & $\U(4)_{\P \T}$ & 4-6 meV \\
$t_A$ & $\U(2)_K \times \U(2)_{K'}$ & 0.5-1 meV 
%\\$\H_{\rm Hund}$ & $\U_C(1) \times \U_V(1) %\times\SU(2)$ & 0.2-0.5 meV 
\\
\hline \hline
\end{tabular}
\caption{Illustration of the $\U(4) \times \U(4)$ symmetry associated with the symmetric part of the interaction $\H_S$. The symmetry corresponds to arbitrary rotations among bands with the same Chern number (top panel). A table illustrating the hierarchy of energy scales and the different symmetries associated with each scale (bottom panel). Here, $\U(4)_{\eta}$ denote the $\U(4)$ subgroup of unitary matrices in $\U(4) \times \U(4)$ commuting with $\eta$.}
\label{SpinfulModelTable}
\end{figure}

Combining the two previous discussions, we find that the interaction in the chiral limit has a  large  $\U(4) \times \U(4)$ symmetry whose generators are $\{t^a, t^a \tau_y \sigma_x, t^a \tau_y \sigma_y, t^a \sigma_z\}$. An intuitive understanding of this result is obtained by observing that in the chiral limit, the form factor $\Lambda_\bq(\bk)$ has the remarkably simple form
\beq
\Lambda_\bq(\bk) = F_\bq(\bk) e^{i \Phi_\bq(\bk) \sigma_z \tau_z}
\label{Lambdaqp}
\eeq
where $F_\bq(\bk)$ and $\Phi_\bq(\bk)$ are two real scalars whose properties are discussed in more detail in the supplemental material. As a result, the interaction is invariant under any unitary rotation which commutes with $\sigma_z \tau_z$ yielding the symmetry $\U(4) \times \U(4)$ corresponding to arbitrary unitary rotations which relate flat-bands  with the same Chern number, as illustrated in Fig.~\ref{SpinfulModelTable}.

\emph{Hierarchy of energy scales}--- In the realistic case where $w_0 \neq 0$ and $\tilde h$ are not negligible, we can estimate the strength of the $\U(4) \times \U(4)$ symmetry breaking by splitting the form factor $\Lambda_\bq(\bk)$ into components $\Lambda^{S/A}_\bq(\bk)$ which commute/anticommute with $R$. Using the remaining symmetries, one can show (supplemental material) that $\Lambda^S_\bq(\bk)$ has the form given in Eq.~\eqref{Lambdaqp}, while $\Lambda^A_\bq(\bk) = \sigma_x \tau_z F^A_\bq(\bk) e^{i \Phi^A_\bq(\bk)\sigma_z \tau_z}$. We can now write the density as $\delta \rho_\bq = \delta \rho^S_\bq + \delta \rho^A_\bq$ with $\delta \rho^{S/A}_\bq$ given by
\beq
\delta \rho^{S/A}_\bq = \sum_\bk \left\{c_\bk^\dagger \Lambda^{S/A}_\bq(\bk) c_{\bk + \bq} - \frac{1}{2} \sum_\bG \delta_{\bG,\bq} \tr \Lambda^{S/A}_\bG(\bk) \right\}
\label{rhopm}
\eeq
We notice that the $R$-symmetric component of the density $\delta \rho^S_\bq$ acts within the same sublattice whereas the $R$ non-symmetric part $\delta \rho^A_\bq$ acts {\em between} sublattices.
This induces a splitting of the interaction into an intrasublattice part $\H_S = \frac{1}{2A} \sum_\bq V_\bq \delta \rho^S_\bq \delta \rho^S_{-\bq}$ which has the full $\U(4) \times \U(4)$ symmetry and an intersublattice part $\H_A = \frac{1}{2A} \sum_\bq V_\bq [\delta \rho^S_\bq \delta \rho^A_{-\bq} + \delta \rho^A_\bq \delta \rho^S_{-\bq} + \delta \rho^A_\bq \delta \rho^A_{-\bq}]$ with only a $\U(4)$ symmetry. Similarly, the form of the dispersion $\tilde h$ is restricted by symmetries to
\beq
\tilde h(\bk) = h_0(\bk) \tau_z + h_x(\bk) \sigma_x + h_y(\bk) \sigma_y \tau_z
\eeq
with the $R$-symmetric (non-symmetric) part given by $h_{x,y}(\bk)$ ($h_0(\bk)$).  Note that, unlike the interaction, the symmetric part acts between sublattices and the non-symmetric part acts within each sublattice). 

Let us denote the typical energy scales associated with $\H_{S}$, $\H_A$, $h_{x,y}(\bk)$ and $h_0(\bk)$ by $U_S$, $U_A$, $t_S$ and $t_A$, respectively (see supplemental material for details). One crucial observation is that even though the realistic value of $w_0/w_1$ is not small, the $R$-breaking terms $U_A$, $t_A$ are smaller by a factor of 3-5 than their $R$-symmetric counterparts $U_S$, $t_S$ as shown numerically in supplemental material and summarized in Fig.~\ref{SpinfulModelTable}. Furthermore, even after accounting for the band renormalization effects, the dispersion $t_S$ is on average smaller by a factor of 3-5 compared to the interaction.

The previous discussion points to a hierarchy of energy scales associated with different symmetries. The largest scale is associated with the intrasublattice interaction $\H_S$ which has the enlarged symmetry $\U(4) \times \U(4)$ implemented by unitary rotations which commute with $\sigma_z \tau_z$. This symmetry is broken at lower energy scales by two different terms. First, the intersublattice  $h_{x,y}$ breaks this down to a single $\U(4)_R$ which commutes with $\sigma_x$ corresponding to the symmetry of the chiral model discussed earlier. Second, the intersublattice interaction $\H_A$ breaks it down to a different $\U(4)_{\P \T}$ subgroup which commutes with $\sigma_x \tau_z$. The presence of both terms thus reduce the symmetry down to $\U(2)_K \times \U(2)_{K'}$ which is the intersection of the two $\U(4)$ subgroups. The intrasublattice dispersion $h_0$ is smaller in magnitude ($\sim$ 0.5-1 meV) and does not break the symmetry any further so it can be neglected. Finally, the intervalley Hund's coupling breaks the symmetry down to $\U_C(1) \times \U_V(1) \times \SU(2)$ at smaller scales. Close to the magic angle all the scales are governed by the interaction, and depend crucially on the structure of the wavefunctions (via $\Lambda_\bq(\bk)$) rather than the detailed $\bq$ dependence of $V_\bq$.

\emph{Energetics and ground state of the spinless model}--- To understand the competition between different states, it is instructive to start by considering the simpler problem of spinless electrons at half filling for which we simply need to replace $\U(4) \to \U(2)$ in the discussion above. Physically, this is equivalent to assuming a spin-unpolarized solution at CN or a spin-polarized solution at half-filling. 

We take the strong coupling limit by assuming that the intrasublattice interaction scale is much larger than the other scales , i.e. $U_S \gg U_A, t_S$, and subsequently solve for the ground states in this limit. For the realistic parameters, $U_S$ is only a factor of 3-5 larger than $U_A$ and $t_S$. However, as we will see, the results of the strong coupling analysis agree remarkably well with the Hartree-Fock numerics, providing an independent justification for the results beyond mean field. We will comment later on the validity of our results for intermediate coupling $U_S \sim t_S$. 

\begin{figure}
    \centering
    \includegraphics[width = 0.45 \textwidth]{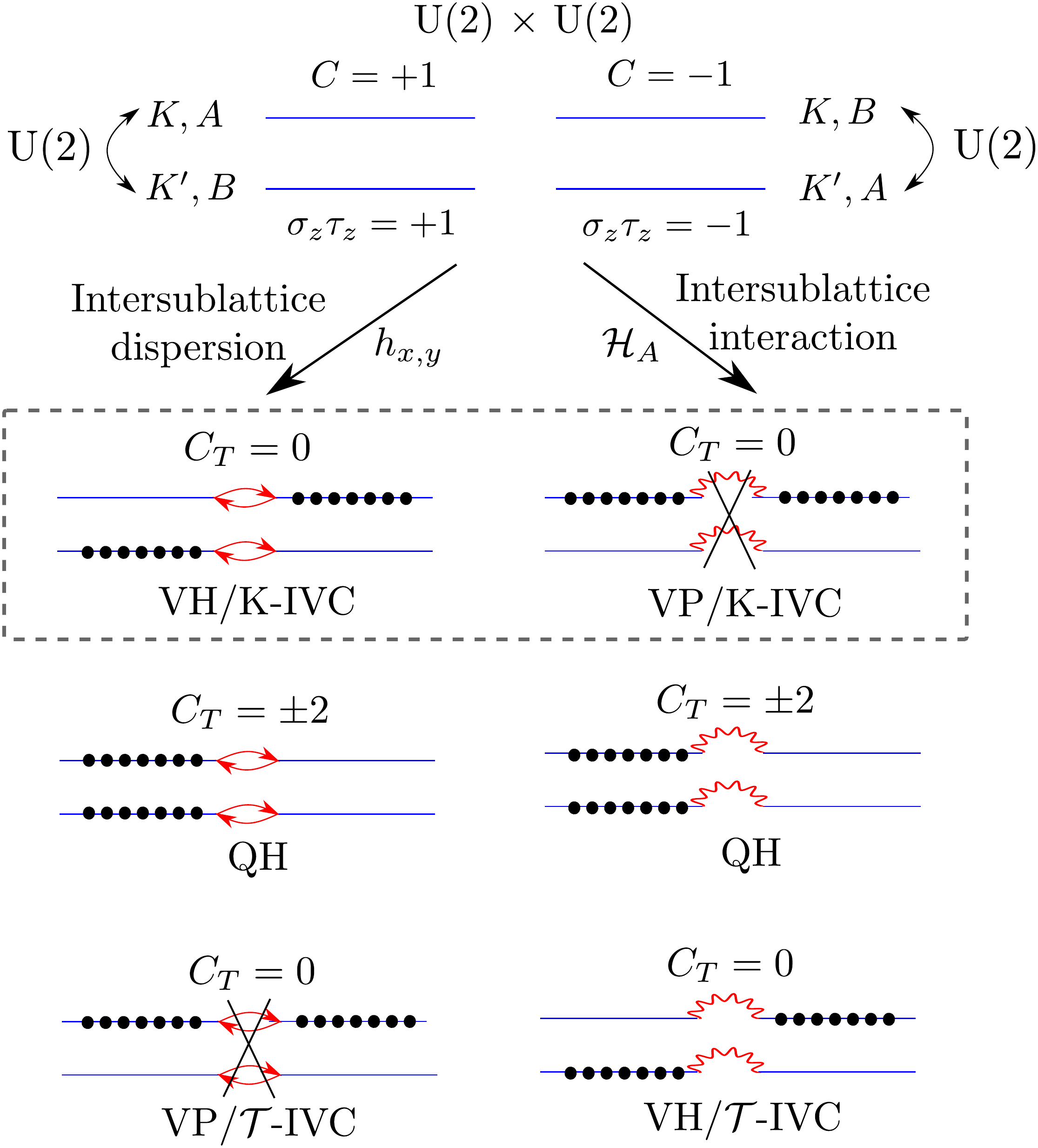} \\
\vspace{0.3 cm}
    \begin{tabular}{c|c|c|c|c}
\hline \hline
Order & Q & Energy & $\eta_{h_{x,y}}$ & $\eta_{\Lambda^A_\bq(\bk)}$ %& Symmetries
\\
\hline
$\T$-IVC & $\sigma_x \tau_+ e^{i \phi} + \text{h.c.}$ & $\lambda$ & $+$ & $-$ %& $e^{i \phi \tau_z} C_2$, $\T$ 
\\
QH & $\sigma_z \tau_z$ & $\lambda - J$ & $-$ & $-$ %& $C_2$, $U_V(1)$ 
\\
VH & $\sigma_z$ & $\lambda - J$ & $-$ & $-$ %& $\T$, $U_V(1)$ 
\\
VP & $\tau_z$ & $0$ & $+$ & $+$ %& $C_2 \T$, $U_V(1)$ 
\\
K-IVC & $\sigma_y \tau_+ e^{i \phi} + \text{h.c.}$ & $-J$ & $-$ & $+$ %& $\tau_z e^{i \phi \tau_z} C_2$, $\tau_z \T$
\\
\hline \hline
\end{tabular}
   \caption{Schematic illustration of the symmetry reduction and ground state selection in the spinless model (top panel). Beginning with the $\u(2) \times \u(2)$ symmetric intrasublattice interaction  $\H_S$, which allows for free rotations within the two $C=1$ and two $C=-1$ levels, the symmetry is lowered by the dispersion $h_{x,y}$ (left) and the intersublattice interaction $\H_A$ (right), which splits the degenerate states. The K-IVC is the unique state which is optimal for both perturbations. Table of the low energy states in the spinless model and how their energy is affected by dispersion $h_{x,y} \propto \sigma_x, \sigma_y \tau_z$ and finite sublattice polarization $\Lambda^A_\bq(\bk) \propto \sigma_x \tau_z, \sigma_y$ (bottom panel). Here, $J \sim t_S^2/U_S$ and $\lambda \sim U_A^2/U_S$ are of the order 1-2 meV and $\eta_x = +/-$ depending on whether the commutator/anticommutator of $Q$ and $x$ vanishes, i.e. $[Q,x]_{\eta_x} = 0$
    }
    \label{SpinlessModel}
\end{figure}

We start by noting that $\H_S$ is a non-negative definite operator for any repulsive interaction $V_\bq > 0$, which implies that any state satisfying $\delta \rho^S_\bq|\Psi \rangle = 0$ for $\bq \neq 0$ is a ground state \cite{Repellin19, Alavirad19, KangVafekPRL}. Next, we note that the diagonal form of $\Lambda^S_\bq(\bk)$ in sublattice and valley implies that $\delta \rho^S_\bq$ annihilates any sublattice or valley ``ferromagnet'' where two of the four sublattice/valley states shown in Fig.~\ref{SpinlessModel} are completely filled. For $\bq$ which is not a reciprocal lattice vector, this follows by noting that the action of $\delta \rho^S_\bq$ changes an electron's momentum by $\bq$ which is impossible in a completely filled or empty band. For reciprocal lattice vector $\bq$, the action of the first term in (\ref{rhopm}) on a completely filled/empty band is finite but cancels exactly against the second term at CN as shown in supplemental material. Simple states satisfying this condition are the QH  $ \sigma_z \tau_z$, VH $\sigma_z$ and VP $\tau_z$ state. More general states are obtained by acting with any $\u(2) \times \u(2)$ rotation which commutes with $\sigma_z \tau_z$ on these simple states yielding a manifold of Slater determinant states labelled by a $\bk$-independent $Q$ satisfying $[Q, \sigma_z \tau_z] = 0$. They fall into two categories: (i) a $\u(2) \times \u(2)$ invariant QH state with a total Chern number $\pm 2$ obtained by filling two bands with the same Chern number and (ii) a manifold of zero Chern number states generated by the action of $\u(2) \times \u(2)$ on the VP state. This manifold includes the VH state as well as two distinct types of IVC orders which break $\u_V(1)$: the Kramers IVC state $\sigma_y \tau_{x,y}$ discussed earlier and a $\T$-symmetric IVC state with $\sigma_x \tau_{x,y}$. Both IVC states hybridize bands with the same Chern number and, as a result, the order parameter can be uniform in $\bk$ and evade the energy penalty due to vortices discussed in earlier works \cite{Bultinck19, Zhang2018, Lee19}.

\begin{figure}
        \centering
        \includegraphics[width=0.48\textwidth]{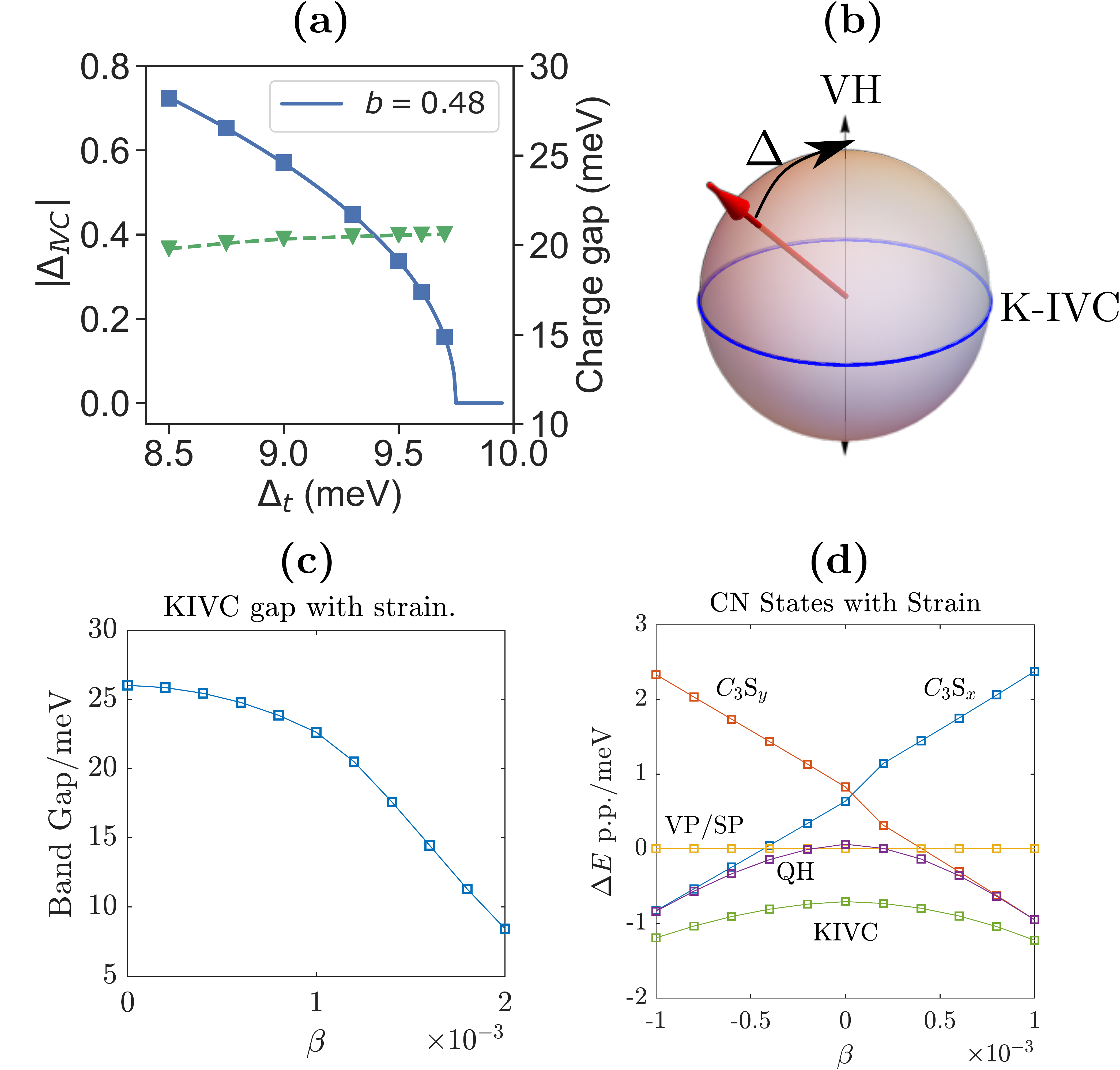}
        \caption{(a) K-IVC order parameter as a function of the sublattice potential $\Delta_t$ on the top layer (blue) fitted to $a(\Delta_{\rm IVC} - \Delta_{t,*})^b$ with $\Delta_{t,*} = 10.35$ meV, with the gap across the transition shown in green. The curve was computed on a $12\times 12$ momentum grid using the parameters $\theta=1.05^o$, $w_0 = 80$ meV, $w_1 = 98$ meV, and $\epsilon = 12$. (b) Schematic illustration of the manifold of low energy states for finite $\Delta$. Since the IVC and VH order parameters anticommute, the order parameter is a vector on $S^2$ which gradually rotates towards the $z$-axis as $\Delta_t$ is increased. (c) IVC gap as a function of the $C_3$-breaking parameter $\beta$ with the energies of the different states as a function of $\beta$ given in (d). %For comparison the results in (c) and (d) were generated using the  subtraction scheme and parameters ($\theta = 1.05^o$, $w_0 = 110$ meV, $w_1 = 82.5$ meV, $\epsilon = 7$) in Ref.~\cite{Liu19} .
        }
        \label{DeltaBeta}
    \end{figure}

Including the dispersion $h_{x,y}(\bk)$ breaks the $\u(2) \times \u(2)$ down to $\u(2)_R$. 
It has the form of an intra-valley, inter-sublattice tunneling with amplitude $h_x(\bk) + i h_y(\bk)$ connecting pairs of opposite Chern bands as shown in Fig.~\ref{SpinlessModel}. Thus, a state in which all pairs of bands connected by $h_{x,y}$ are either both full or both empty is annihilated by $h_{x,y}$ since the tunneling processes are completely blocked. This is equivalent to $[Q, \sigma_x] = 0$. This can be seen by noting that commutation with both $\sigma_x$ and $\sigma_z \tau_z$ means that $Q$ is proportional to the identity in the $\SU(2)$ pseudo-spin variable $(\sigma_x, \sigma_y \tau_z, \sigma_z \tau_z)$ whose $z$-component is the Chern number and $x$, $y$ components correspond to the tunneling $h_{x,y}$, i.e $Q$ describes to a state with zero total pseudo-spin which is annihilated by the pseudo-spin flip operators $\propto h_{x,y}$.
For the remaining states, the action of $h_{x,y}$ creates an electron-hole (e-h) excitation between these pairs of bands. 
Since the electron and hole carry opposite Chern numbers, the electron-hole excitations always have a finite energy of the same order as $U_S$ as shown in the supplemental material. This can be understood by noting that the condensation of such electron-hole pairs is equivalent after a particle-hole transformation to superconducting pairing in a $\pm 2$ Chern band which is known to be energetically unfavorable \cite{Bultinck19}. The energy due the tunneling $h_{x,y}$ can be computed within second order perturbation theory leading to an energy reduction $J \sim t_S^2/U_S \sim $ 1-2 meV. This gain, which resembles antiferromagnetic "superexchange", is due to virtual tunneling processes between pairs of bands connected by $h_{x,y}$ which is maximized if only one band is filled in each pair. This is equivalent to the condition $\{Q,\sigma_x\}$ which is satisfied by two types of states:(i) a $\u(2)$-invariant QH state with Chern number $\pm 2$ and (ii) a manifold of states with vanishing Chern number isomorphic to $\u(2)/\u(1) \times \u(1) \simeq S^2$ generated by the VH and K-IVC states which form a sphere (see Figure \ref{DeltaBeta}b).

The intersublattice part of the interaction $\H_A$ breaks $\u(2) \times \u(2)$ to a different $\u(2)_{\P \T}$ subgroup. Because the cross-terms $\delta \rho_\bq^S \delta \rho_{-\bq}^A + h.c.$  in $\H_A$ are already guaranteed to vanish on the ground-state manifold of $\H_S$, and the residual $\delta \rho_\bq^A \delta \rho_{-\bq}^A$ is positive definite, $\H_A$ selects the submanifold of ground states annihilated by $\delta \rho^A_\bq$. Due to the structure of the intervalley form factor $\Lambda^A_\bq(\bk) \propto \sigma_x \tau_z, \sigma_y$, these states satisfy the condition $[Q, \sigma_x \tau_z] = 0$ forming the manifold $\u(2) /\u(1) \times \u(1) \simeq S^2$ generated by the VP and K-IVC. The energies of the other states is increased by an amount of the order $\lambda \sim U_A^2/U_S \sim $ 1 meV (see supplemental material). 

Thus, in the presence of both $h_{xy}$ and $\H_A$, the K-IVC, which benefits from both perturbations, has the lowest energy followed by the VP and QH/VH (the latter two are degenerate) whose competition is determined by the relative strength of the intersublattice interaction $U_A^2/U_S$ and the energy reduction due to superexchange $t_S^2/U_S$. This is consistent with the numerical results in Fig.~\ref{HFnumerics}, where the energies of the VP state and the QH/VH state cross as a function of $w_1$ which controls both $h_{x,y}$ and $\H_A$. At a fixed $w_1$, decreasing $w_0$ whose main effect is decreasing $\H_A$ clearly favors the VH/QH states and makes them closer in energy to the K-IVC ground state. The $\T$-IVC state, which was not seen in the numerics, is disfavored by both and has the highest energy.

In the realistic magic angle parameter regime, the dispersion scale $t_S$ is only a factor of 3-5 smaller than the interaction scale $U_S$ and some states may become energetically competitive by optimizing this part first. Indeed, this eventually occurs away from the magic-angle when the dispersion becomes comparable to the interaction scale. The simplest such states are semimetallic (SM) solutions preserving both $C_2 \T$ and $U_V(1)$ \cite{Liu19}, which are characterized by 
\beq
Q_{\rm SM}(\bk) = \sigma_x e^{i \phi(\bk) \sigma_z \tau_z}
\eeq
away from the isolated $\bk$ points at which the gap vanishes where the phase $\phi(\bk)$ winds by $\pm 2\pi$. Such SM states also break $C_3$ for realistic values of the parameters $w_0$ and $w_1$ \cite{Liu19}.  Due to the topology of the bands, the phase $\phi(\bk)$ winds twice around the Brillouin zone which means it has at least two vortices (this assumes a smooth gauge choice). Another way to see this is by noting that this order parameter can be obtained by condensing electron-hole pairs discussed earlier, thus gaining energetically from the dispersion but paying an energy penalty $\sim U_S$. In fact, at any finite value of $t_S$, the insulating order parameters corresponding to QH, VH or K-IVC (those benefiting from the "antiferromagnetic" coupling) develop a small component $\sim t_S/U_S$ parallel to $Q_{\rm SM}$ since the corresponding order parameters anticommute. The SM component grows with increasing $t_S$, which results in a gradual reduction of the gap until $t_S \sim U_S$ where the insulating phase disappears \cite{Liu19}. This has important implications for the effect of strain on the insulating state as we discuss later. 

\emph{Charge neutrality: Ground state and spin structure --} Upon including spin, we can similarly study the manifold of ground states at CN starting with the states minimizing the intrasublattice interaction $\H_S$ which satisfy $[Q,\sigma_z \tau_z]$. These are obtained by completely filling 4 of the 8 bands in Fig.~\ref{SpinfulModelTable}. $h_{xy}$ selects states satisfying $\{Q, \sigma_x \} = 0$. These states can be divided into three classes: (i) a spin-unpolarized QH state with Chern number $\pm 4$ obtained by filling all 4 bands with the same Chern number, (ii) a manifold of states with Chern number $\pm 2$ obtained by filling 3 bands with the same Chern number and one band with opposite Chern number, and (iii) a manifold of states obtained by filling 2 bands in each Chern number sector. The states in (ii) are mixed states corresponding, for instance, to a QH state in one spin species and a VH or IVC state in the other and they form the manifold $\U(4)/\U(3) \times \U(1)$. States in (iii) include the spin-unpolarized versions of the spinless phases discussed earlier including the VH  and K-IVC states, which form the manifold $\U(4)/\U(2) \times \U(2)$. In contrast, the interaction $\H_A$ selects states satisfying $[Q, \sigma_x \tau_z] = 0$ which include spin/valley polarized states as well as spin-unpolarized K-IVC states. However, the spin/valley polarized states do not benefit from the dispersion. Thus, combining the effect of the dispersion and $\H_A$ we are left with K-IVC as the unique state that is maximally stabilized by both perturbations.

Note that the spin-unpolarized K-IVC state is not invariant under the action of $\U(2)_K \times \U(2)_{K'}$ rotations. Instead, this action generates a manifold of states which are degenerate with respect to $\H_{\rm eff}$. This manifold can be parameterized by a single 2$\times$2 unitary matrix $V$ in spin space with $Q = \sigma_y (\tau_+ V + \tau_- V^\dagger)$,  $\tau_\pm = \frac{1}{2} (\tau_x \pm i \tau_y)$. To understand the structure of these states, we write the manifold as $\U(2) \simeq \U(1) \times \SU(2)$ which can be parametrized as $V = e^{i \phi} e^{i \frac{\theta}{2} \bn \cdot \bs}$. Thus, a given K-IVC state is specified by choosing a spin quantization axis $\bn$ on $S^2$ and specifying two $\U(1)$ K-IVC phases $\phi \pm \frac{\theta}{2}$ for the up and down spins along $\bn$. Note however that the spin axis $\bn$ loses meaning for the spin-singlet state $\theta=0$. The intervalley-Hunds coupling fixes the value of the relative phase $\theta$ between the K-IVC states for up and down spins. An antiferromagnetic coupling, perhaps driven by phonons \cite{Chatterjee19}, leads to  $\theta=0$. As expected this is the spin singlet K-IVC state, where the orbital currents from opposite spins add. On the other hand, ferromagnetic Hunds coupling leads to$\theta = \pi$, i.e. a spin `triplet' K-IVC state. At this special value, the orbital currents of the oppositely directed spins cancel, leaving behind circulating spin currents (see Figure \ref{fig:KIVC}).

\emph{Half Filling: Ground State and Spin Structure--} While we have largely focused on charge neutrality $\nu=0$, let us now briefly discuss half filling i.e. $\nu=\pm 2$, leaving a more through discussion for the future. At half-filling $\nu = -2$ (the case of $\nu = 2$ can be deduced by performing a particle hole transformation on the conclusions below), the ground states of $\H_S$ are obtained by filling 2  out of the 8 bands encoded by the condition $[Q,\sigma_z \tau_z] = 0$. In contrast to CN, these states are not completely annihilated by the operator $\delta \rho^S_\bG$ for reciprocal lattice vectors $\bG$. Instead, the action of $\H_S$ on these states yields a constant energy that does not affect their energy competition. However, such contribution may affect the competition between the $\nu = \pm 2$ insulating states and metallic or  superconducting phases emerging from the $\nu = 0$ state. We leave investigating such competition to future works. Within the manifold of groundstates of $\H_S$, states can gain energetically from tunneling if at most one out of each pair of bands coupled through $h_{x,y}$ is filled. The resulting states either have (i) Chern number $\pm 2$ such as valley and sublattice polarized or spin-polarized QH states (forming the manifold $\U(4)/\U(2) \times \U(2)$) or (ii) Chern number 0 such as the spin-polarized VH or K-IVC states (forming the manifold $\U(4)/\U(2) \times \U(1) \times \U(1)$). Again, the interaction $\H_A$ selects instead states satisfying $[Q,\sigma_x \tau_z] = 0$ which include spin and valley polarized states and spin-polarized K-IVC. The ground state manifold in the presence of both band dispersion and $\H_A$ is the K-IVC state. The set of nearly degenerate K-IVC states is obtained by acting with $\U_K(2) \times \U_{K'}(2)$ on the spin-polarized K-IVC state. The resulting manifold is isomorphic to $\U(1) \times S^2 \times S^2$ denoting the K-IVC phase and the direction of the spin in each valley which can be chosen independently. Intervalley Hund's coupling locks the spin in the two valleys to be either parallel ($J<0$ ferromagnetic Hunds coupling) or anti-parallel ($J>0$ antiferromagnetic Hund's coupling). In both cases  spatially varying orbital magnetization currents are present. A full Hartree-Fock numerical analysis of this case is left to future work but it is worth noting that band renormalization effects at half-filling are expected to be larger than at CN, resulting in smaller gaps. 

\emph{Phenomenology of the K-IVC}---
We now comment on the phenomenological consequences of the K-IVC order:

\begin{itemize}
    \item{Circulating currents.}
        Fixing a spin species,  the lattice-scale current  $j_{ij}$ in the K-IVC ground state manifests a pattern of circulating currents which triples the unit cell, as shown in Fig.~\ref{fig:KIVC}.
        The typical current (or equivalently the typical magnetization density) is of the order of Microamperes i.e. $j \sim \mu \textrm{A}$.
        This finding is consistent with the estimate $j \sim e \frac{v_F}{a} \left(\frac{a}{L_M} \right)^2 \sim 0.7 \mu \textrm{A}$ obtained by assuming each electron in the flat band is circulating at velocity $v_F$.
        In the spin-singlet K-IVC, the two spin-species carry the same current, and the state is thus an orbital-magnetization density wave.
        The spin-triplet K-IVC $Q = \mathbf{n}\cdot\mathbf{s} \, \tau_{x/y} \sigma_y$, however, is invariant under the usual \emph{spinful} time-reversal operation $\textrm{TR} = i s_y \tau_x K$. Hence the two spin-species carry opposite current and the magnetization cancels - instead, there are circulating \emph{spin} currents.
        
        Nevertheless, both cases triple the unit cell. 
        In the presence of  umklapp scattering, this tripling will manifest as small bond distortions or topographic changes reminiscent of a Kekule pattern, which may be observable in atomically-resolved STM spectroscopy.
     
    \item{Landau fan.} Due the $\mathcal{T}'$ Kramers degeneracy, the conduction (valence) bands  of the K-IVC (Fig.~\ref{HFbands}) have a doubly degenerate band minimum (maxima) at the mini-$\Gamma$ point. Per spin, they consist of a pair of bands, which we label $Z = \pm 1$, which disperse quadratically.  Both bands carry trivial $C_3$ quantum number, and thus to leading order within a $k.p$ approach the Hamiltonian for the conduction band-minima is
    \begin{align}
        H_{\Gamma} = \frac{ (\mathbf{p} - \mathbf{A})^2}{2 m^\ast} + B (m_\Gamma \hat{Z} +  g_s \mu_B \hbar \frac{s_z}{2}) +  \mathcal{O}(p^3)
    \end{align}
    where $m^\ast$ is the effective mass, $\nabla \times \mathbf{A} = B$ is the external magnetic field,  $m_\Gamma$ is the orbital magnetization of the bands at the $\Gamma$-point (which is odd under $\mathcal{T}'$), and $g_s$ is the $g$-factor for spin.
    The low-field Landau-level spectrum is thus $\epsilon_N =  B ( \frac{\hbar e}{m^\ast} (N + \frac{1}{2}) + m_\Gamma Z + g_s \mu_B \hbar \frac{s_z}{2}) + \cdots$, with an analogous result for the valence band.
    Neglecting $g_s$ and the magnetization $m_\Gamma$, the Landau-fan would thus have a $\nu = \pm 0, 4, 8, \cdots$ degeneracy arising from spin and  $\mathcal{T}'$-Kramers degeneracy.  With $m_\Gamma$, however, this degeneracy splits, $\nu = \pm 0, 2, 4, \cdots$, with the relative strength of the splitting depending on the ratio of $\hbar \frac{e}{ m^\ast}$ to $m_\Gamma$. Experiments reporting a charge-gap at neutrality find oscillations at $\nu = \pm 0, 2, 4, 8, \cdots$ \cite{efetov}, which seemingly combines the two. This may be because at higher $N$ or $B$, the $\mathcal{O}(p^3)$ terms become important. Also, one important caveat is that we find the K-IVC band structure around the $\Gamma$ point to be sensitive to the twist angle, so the above analysis may not always apply. A full quantitative calculation of the quantum oscillations therefore remains as a useful direction for future work.

    \item{$\mathbb{Z}_2$-topology.}  Remarkably, when restricting to a spin-species, the K-IVC  is a topological insulator protected by Kramers time-reversal $\T'$ and $\U(1)$ charge conservation. This is expected since it consists of two IVCs with opposite Chern number ($|KA\rangle + |K'B\rangle$ and $|KB\rangle + |K'A\rangle$) related by  $\T'$. Note however this does not automatically imply edge states since the fractional translation $\tau_z$ involved in $\T'$ may be broken by a rough edge.

    \item{Phase-transitions.} Finally, on breaking various symmetries the K-IVC can be weakened or destroyed as discussed below.

\end{itemize}

\emph{Effect of single-particle perturbations}--- Due to the presence of an enlarged $\U(4) \times \U(4)$ symmetry which is only broken by relatively small terms which settle the energy competition among a few low energy states, we expect the ground state to be sensitive to symmetry lowering perturbations such as sublattice potential, strain and magnetic field. The presence of a sublattice potential $\Delta \sigma_z$ is associated with alignment with hBN substrate which explicitly favors the VH state ($Q = \sigma_z$) over the K-IVC state. Assuming a fixed spin structure ($Q \propto s_0$ or $\bn \cdot \bs$), the two order parameters anticommute forming an O(3) vector living on $S^2$ as shown in Fig.~\ref{DeltaBeta}. As $\Delta$ is increased, this vector rotates towards the $z$-axis (VH) until it points completely along the $z$-direction restoring $\U_V(1)$ symmetry as shown in Fig.~\ref{DeltaBeta}. As a result, we do not expect this phase transition to be associated with a gap closing in the fermionic sector which is verified numerically in Fig.~\ref{DeltaBeta}.

Next, we consider the effect of strain which influences the non-interacting band structure in two distinct ways \cite{BiFuStrain}. First, it renormalizes the bandwidth leading to an increase in the magnitude of the single particle dispersion $t_S$. As discussed earlier, this will favor the semimetallic solution and has the effect of gradually reducing the gap in the K-IVC solution by increasing the SM component. The second effect of strain is the explicit breaking of $C_3$ symmetry. This can be taken into account phenomenologically following Refs.~\cite{ZhangPo, Liu19} by rescaling one of the Moir\'e hopping parameters by $1+\beta$. This introduces explicit $C_3$ symmetry breaking in the dispersion $h_{x,y}$ resulting in a linear coupling to the energy of the $C_3$-breaking SM as shown in Fig.~\ref{DeltaBeta}d. The VH and K-IVC states will respond to $\beta$ by increasing their SM component leading to a {\it quadratic} decrease of the VH and K-IVC energies and gaps as a function of $\beta$ seen in Fig.~\ref{DeltaBeta}. With increasing $\beta$, the energy of the three orders approach each other whereas other states such as VP are not affected. %Since the creation of electron-hole pairs is associated with a gapless order parameter due to vortices, the gap of the VH and IVC states goes down quadratically with $\beta$ as shown in Fig.~\ref{DeltaBeta}. 
It is worth noting that semimetallic behavior in transport can also emerge purely from disorder, even when the ground state of the clean system is insulating \cite{Thomson19}.

Finally, let us comment briefly on the effect of magnetic field. The Zeeman coupling depends on the spin structure and its effect on the gap depends non-trivially on the type of low-lying excitations \cite{Chatterjee19}. On the other hand, the orbital effect of the magnetic field can be understood as follows. For in-plane field, its main effect is to break $C_3$ symmetry, shifting the Dirac points away from the Moir\'e K and K$'$ points. In this regard, the effect is similar to the $C_3$-breaking perturbation discussed above yielding a quadratic decrease of the gap with in-plane field which is consistent with the observation of Ref.~\cite{Dean-Young}. On the other hand, an out-of-plane field is associated with a relatively large Chern-Zeeman effect $\sim \sigma_z \tau_z$ which shifts the energies of the opposite Chern bands relative to each other. As a result, it is expected to drive a transition to a QH state with Chern number $\pm 4$ at neutrality and $\pm 2$ at half-filling. We leave a more quantitative discussion for the effect of magnetic field to future works.

\emph{Conclusions}--- To summarize, based on both numerical and analytical arguments, we propose that the insulating state observed at charge neutrality in pristine MATBLG \cite{efetov} is the K-IVC state, i.e. an inter-valley coherent state with an emergent spinless Kramers time-reversal symmetry $\mathcal{T}'$. Interestingly, modulo spin degeneracy, the K-IVC is a non-trivial topological insulator protected by $\mathcal{T}'$. As a result, it does not admit a real space strong coupling "Mott" description as long as the locality of time-reversal and valley $\U(1)$ symmetries is preserved. This, in turn, suggests that the momentum space description employed here which closely parallels multilayer quantum Hall problems is more suited to MATBG than real space descriptions \cite{Thomson18, IsobeFu, KangVafeKPRX, Seo19}, at least when restricted to the space of flat bands at integer fillings. It is worth noting that despite some similarities to a previously proposed intervalley-coherent order \cite{KangVafekPRL}, our state differs in several crucial aspects, such as the absence of time-reversal symmetry and the presence of non-trivial band topology which forbids a localized Mott description. Spontaneous magnetization density wave states have been discussed in other settings notably in the context of the cuprates as the staggered flux \cite{WenLeeReview}  and d-density wave states \cite{d-densitywave} and loop current states\cite{VarmaLoopCurrent} (for a recent discussion of loop current states motivated by tBG, see Ref. \cite{Nandkishore}), and in untwisted bilayer graphene \cite{Varma-bilayer,Venderbos}. While reminiscent of the state discussed here, an important difference is that the  K-IVC is very weakly coupled to the underlying lattice. Thus the spontaneously breaking of the enlarged U(1)$_{\rm valley}$ symmetry leads to new consequences including gapless Goldstone modes and emergent Kramers time reversal symmetry.

One important issue that is worth highlighting is that we do not expect a finite temperature phase transition into the K-IVC state, \emph{even though it breaks the discrete time-reversal symmetry $\T$}. The reason is that the time-reversal symmetry breaking is non-trivially intertwined with the breaking of the continuous valley charge conservation symmetry. This can be seen by noting that the presence of the Kramers time-reversal symmetry $\mathcal{T}'=\tau_y K$ implies that there is no order parameter with non-vanishing expectation value in the K-IVC state which breaks $\T$ without breaking $\U(1)$ valley charge conservation.

The analytical arguments in favor of the K-IVC state are based on the presence of an approximate $\U(4) \times \U(4)$ symmetry. One consequence of this approximate symmetry is that small perturbations to the BM band spectrum coming from e.g. h-BN alignment or strain can destroy the K-IVC state and instead give rise to a valley-Hall or semi-metallic state at charge neutrality. It is therefore important to have an estimate of the magnitude of these effects in different devices. Our analysis has a natural generalization to doped systems with two additional electrons or holes per Moir\'e unit cell ($\nu = \pm 2$), so we expect a spin-polarized version of the K-IVC state to occur at those fillings. At odd integer fillings the situation is different. Applying our construction to odd filling inevitably leads to anomalous Hall insulators, which is at odds with the present experimental data in tBG devices which are unaligned with the h-BN substrate. In fact, our analysis points to the possibility of different types of states at odd filling since, unlike the K-IVC states at even filling, no  translationally symmetric Slater determinant state takes advantage of all the terms in the Hamiltonian. In addition, band renormalization effects are expected to play a bigger role, particularly at $\nu = \pm 3$ where mixing with remote bands is more likely \cite{Xie2018}.

The K-IVC state exhibits a very subtle type of symmetry-breaking order, leading to an interesting phenomenology. Depending on the spin texture of the K-IVC state, which is only determined by the small intervalley Hunds terms, we have put forward a physical interpretation of the K-IVC state as either an `orbital-magnetization density wave' on the atomic scale, or a state with circulating spin currents. These types of order are presumably hard to directly detect experimentally, but leave their imprint on the electronic structure. Proposals for a smoking-gun experiment to identify the K-IVC state is left to future work. 

Finally, let us comment briefly on the implications of our findings for superconductivity. The presence of the Kramers time-reversal symmetry $\mathcal{T}'$ has  important implications for the nature of superconducting states which are proximate to the K-IVC order. Recall that in conventional superconductors with spin orbit coupling  the Anderson theorem \cite{Anderson} protects pairing between Kramers time-reversal partners, even in the presence of non-magnetic impurities. Similarly, superconductivity is expected to remain robust in the presence of K-IVC order, as long as electrons related by the $\mathcal{T}'$ symmetry are being uniformly paired.  The K-IVC mean-field band structure indicates  that small electron or hole doping will lead to concentric Fermi surfaces around the $\Gamma$ point, which are related to one another by $\mathcal{T}'$ symmetry. Hence a Fermi surface coexisting with K-IVC order can  be destabilized by coupling to phonons and/or order parameter fluctuations giving rise to the superconducting state. We leave a more detailed analysis of the nature of the superconducting states and their connection to the K-IVC for future work.

\emph{Acknowledgement.}--- We thank T.~Senthil for helpful discussions. This work was partly supported by the Simons Collaboration on Ultra-Quantum Matter, which is a grant from the Simons Foundation (651440, A.~V.) and A.~V., E.~K., S.~L. were supported by a Simons Investigator grant. Research at Harvard is partially supported as part of the Center for the Advancement of Topological Semimetals, an Energy Frontier Research Center funded by the U.S. Department of Energy (DOE), Office of Science, Basic Energy Sciences (BES) through the Ames Laboratory under its Contract No. DE-AC02-07CH11358. E. K. was supported by the German National Academy of Sciences Leopoldina through grant LPDS 2018-02 Leopoldina fellowship. S.~C. acknowledges support from the ERC synergy grant UQUAM. M.~Z. and N.~B. were supported by the DOE, office of Basic Energy Sciences under contract no. DE-AC02-05-CH11231.

\bibliography{refs}

\pagebreak
\widetext
\begin{center}
\textbf{\large SUPPLEMENTAL MATERIAL:\\ Ground State and Hidden  Symmetry of Magic Angle Graphene at Even Integer Filling}
\end{center}
%%%%%%%%%% Merge with supplemental materials %%%%%%%%%%
%%%%%%%%%% Prefix a "S" to all equations, figures, tables and reset the counter %%%%%%%%%%
\setcounter{equation}{0}
\setcounter{figure}{0}
\setcounter{table}{0}
\setcounter{page}{1}
\makeatletter
\renewcommand{\theequation}{S\arabic{equation}}
\renewcommand{\thefigure}{S\arabic{figure}}
\renewcommand{\bibnumfmt}[1]{[S#1]}

\tableofcontents

\section{Hamiltonian}
\subsection{Bistritzer-Macdonald model}
Our starting point is the Bistritzer-MacDonald (BM) model of the TBG band structure \cite{MacDonald2011}, which we now briefly review. We begin with two layers of perfectly aligned (AA stacking) graphene sheets extended along the $xy$ plane, and we choose the frame orientation such that the $y$-axis is parallel to some of the honeycomb lattice bonds. Now we choose an arbitrary atomic site and twist the top and bottom layers around that site by the counterclockwise angles $\theta/2$ and $-\theta/2$ (say $\theta>0$), respectively. When $\theta$ is very small, the lattice form a Moir\'e pattern with very large translation vectors; correspondingly, the Moir\'e Brillouin zone (MBZ) is very small compared to the monolayer graphene Brillouin zone (BZ). In this case, coupling between the two valleys can be neglected. If we focus on one of the two valleys, say $K$, then the effective Hamiltonian is given by: 
\begin{align}\label{BMHam}
\H_S= \sum_l \sum_{\bk} f^\dagger_{l}(\bk) h_{\bk}(l\theta/2) f_{l}(\bk) + \left(\sum_{\bk}\sum_{j=1}^3 f^\dagger_{t} (\bk+\bq_j) T_j f_{b}(\bk) + h.c.\right). \end{align}
Here, $l=t/b$ is the layer index, and $f_l(\bk)$ is the $K$-valley electron originated from layer $l$. The sublattice index $\sigma$ is suppressed, thus each $f_l(\bk)$ operator is in fact a two-column vector. In the original BM model, $h_{\bk}(\theta)$ is the linearized monolayer graphene $K$-valley Hamiltonian with twist angle $\theta$: 
\beq\label{linear}
	h_{\bk}(\theta)=\hbar v_F (k_x \sigma_x + k_y \sigma_y) e^{-i \theta \sigma_z}
\eeq
where $v_F$ is the Fermi velocity. In our numerics, we do not use the linearized dispersion \eqref{linear}, but we replace it with the complete mono-layer graphene Hamiltonian. This means that $h_{\bk}(\theta)$ is instead given by
\beq
h_{\bk}(\theta) = h_{MLG}(\bK+R(\theta)\bk) = \left(\begin{matrix} 0 & g_0(\bK+R(\theta)\bk) \\ g_0^*(\bK+R(\theta)\bk) & 0 \end{matrix}\right)\, ,
\eeq
where $R(\theta)$ is the two-dimensional rotation matrix which rotates over an angle $\theta$, $g_0(\bk)$ is given by
\beq
g_0(\bk) = -t_0\sum_{l=1}^3 e^{i\bk\cdot \bdelta_l}\, ,
\eeq
and $\bdelta_l$ are the three vectors connecting an $A$-sublattice site to its neighboring $B$-sublattice sites.

To define the second term in the BM Hamiltonian \eqref{BMHam} describing the inter-layer tunneling, we write the $K$-vector of layer $l$ as $K_l$. With this notation, $\bq_1$ is defined as $K_b-K_t$. $\bq_2=O_3\bq_1$ is the counterclockwise $120^\circ$ rotation of $\bq_1$, and $\bq_3=O_3\bq_2$. Finally, the three matrices $T_i$ are given by 
\beq
	T_j = w_0 \sigma_0 + w_1 \sigma_x e^{\frac{2\pi i}{3} (j-1) \sigma_z} 	
	\label{T123}
\eeq
Unless otherwise stated, we will use the values
\beq
t_0 = 2.8 \text{ eV}, \qquad w_0 = 80 \text{ meV}, \qquad w_1 = 110 \text{ meV}, \qquad \theta = 1.05^o
\label{parameters}
\eeq

The single particle Hamiltonian within each valley $\H_\pm$ is invariant under the following symmetries
\beq
 C_3 f_{\bk} C_3^{-1} = e^{-\frac{2\pi}{3} i \tau_z \sigma_z} f_{C_3 \bk}, \quad (C_2 \T) f_\bk (C_2 \T)^{-1} = \sigma_x f_\bk, \qquad \M_y f_\bk M_y^{-1} = \sigma_x \mu_x f_{M_y \bk},
\eeq
where $M_y \bk = (k_x, -k_y)$. In addition, the two valleys are related by time-reversal symmetry given by
\beq
\T f_\bk \T^{-1} = \tau_x f_{-\bk}.
\eeq
Here, $\boldsymbol\sigma, \boldsymbol\tau$ and $\boldsymbol \mu$ denote the Pauli matrices in sublattice, valley and layer spaces, respectively. As a result, we can also write $C_2$ as
\beq
C_2 f_\bk \C_2^{-1} = \tau_x \sigma_x f_{-\bk}
\eeq

In addition, at small angles, we can neglect the $\theta$ dependence of $h_\bk(\theta)$. In this case, we have the extra unitary particle-hole symmetry given by
\beq
\P f_\bk \P^{-1} = i \sigma_x \mu_y f^\dagger_{-\bk}
\label{PH}
\eeq
In addition, we have $U_V(1)$ valley charge conservation given by
\beq
U_V f_\bk U_V^{-1} = e^{i \phi \tau_z} f_\bk
\eeq

For the first-quantized Hamiltonian, the symmetries can be written as illustrated in Table \ref{Symmetries}. Here, we made the replacement $\M \rightarrow i \M$ whenever necessary to make all $\Z_2$ unitary symmetries square to $+1$ and $[\T, \P] = 0$.

\subsection{Chiral model}
The chiral limit corresponds to taking the limit of vanishing intrasublattice Moire hopping $w_0 = 0$. In this case, the Hamiltonian has the extra anti-unitary chiral symmetry
\beq
\S f_\bk \S^{-1} = \sigma_z f^\dagger_\bk
\eeq
In this case, we can perform the gauge transformation $f_{l,\bk} \mapsto e^{\frac{i}{2} l \theta \sigma_z} f_{l,\bk}$ to get rid of the $\theta$ dependence in the first term in the Hamiltonian. As a result, the particle-hole symmetry (\ref{PH}) is exact at all angles. This means that we can combine $\S$, $\T$ and $\P$ to get a $\Z_2$ unitary symmetry $R = \S \P \T$
\beq
R f_\bk R^{-1} = \tau_x \sigma_y \mu_y f_\bk
\eeq 
which flips layer, valley and sublattice. Combining this symmetry with the different symmetries of the model, we can generate different versions of the symmetries, for example new time-reversal symmetry $\T' = \T U$ acting as
\beq
\T' f_\bk {\T'}^{-1} = \sigma_y \mu_y f_{-\bk}
\eeq
This time-reversal symmetry flips layer and sublattice indices but acts within the same valley. We can also define a new $C_2$ symmetry $C'_2 = C_2 U$ which leaves valley and sublattice index invariant
\beq
C'_2 f_\bk {C'_2}^{-1} = \sigma_z \mu_y f_{-\bk}
\eeq
In addition, we can combine $\T'$ with $C_2$ to get a new $C_2 \T$ symmetry which leaves momentum invariant but interchanges valley and sublattice
\beq
(C_2 \T') f_\bk (C_2 \T')^{-1} = i \tau_x \mu_y \sigma_z f_\bk
\eeq
which satisfies $(C_2 \T')^2 = -1$.

\begin{table}
\center
\begin{tabular}{c|c|c|c|c|c|c|c}
\hline\hline
 & \multicolumn{5}{c|}{ $w_0 \neq 0$} & \multicolumn{2}{c}{$w_0 = 0$}\\
\hline\hline
& $\T$ & $C_2$ & $C_2 \T$ & $\P$ & $\P \T$ & $\S$ & $R = \S \P \T$  \\
\hline
Original basis & $\tau_x \K$ & $\tau_x \sigma_x$ & $\sigma_x \K$ & $i \sigma_x \mu_y \K$ & $\tau_x \sigma_x \mu_y$ & $\sigma_z$ & $\tau_x \sigma_y \mu_y$ \\ 
\hline
Sublattice basis & $\tau_x \K$ & $\sigma_x \tau_x e^{i \theta(\bk)}$ & $e^{i \theta(\bk)} \sigma_x \K$ & $\tau_z \sigma_y  \K$ & $\tau_y \sigma_y$ 
& $\sigma_z$ & $\tau_y \sigma_x$  \\  
\hline\hline
\end{tabular}
\caption{Symmetries of the BM model with non-vanishing/vanishing intrasublattice hopping in the original microscopic basis and the projected sublattice basis. In the latter, the gauge is chosen such that $\T = \tau_x \K$ and $P = \tau_z \sigma_y \K$ and $\K$. Here, $\tau$, $\sigma$, $\mu$, $\gamma$ denote the Pauli matrices in the valley, sublattice, layer and band, respectively.
}
\label{Symmetries}
\end{table}

\subsection{Interaction and projection onto active bands}
In the following, we derive the form of the interaction when projecting onto a set of active bands. Let $c^\dagger_{\alpha}(\bk)$ be the creation operator for the energy eigenstate labelled by the combined index $\alpha = (\mu, n)$ which includes the flavor index $\mu$ labelled by spin $s=\uparrow,\downarrow$ and valley $\tau=\pm$ and band index $n$. The fermion creation operator $f^\dagger_{\mu,a}(\br)$, with $a$ denoting the layer and sublattice indices $a= (l,\sigma)$, in the continuum model is defined by expanding the graphene lattice fermion creation operator close to the K and K' points as $f^\dagger_{s,\sigma,l}(\br) = e^{i \bK_l\cdot \br}f^\dagger_{(s,\tau=+),a}(\br) + e^{-i \bK_l\cdot \br} f^\dagger_{(s,\tau=-),a}(\br)$. $f^\dagger_{\mu,a}(\bq)$ is its Fourier transform in terms of the continuous momentum $\bq$ which is not restricted within the first Moir\'e Brillouin zone. $c^\dagger$ and $f^\dagger$ are related to each other by the $k$-space wave functions as follows: 
\begin{align}
	c^\dagger_{\mu,n}(\bk)=\sum_{\bG,a} u_{\tau,n;\bG,a}(\bk) f^\dagger_{\mu,a}(\bk + \bG), 
\end{align}
where $\bG$ is a Moir\'e reciprocal lattice vector and we used the fact that the wave functions are spin-independent. 
Once we choose a gauge of $u_{\tau,n;\bG,a}(\bk)$ for all $\bk$ in some MBZ, $c^\dagger(\bk)$ are defined in terms of the $f^\dagger(\bq)$ for those $\bk$. Due to the band topology, it is generally impossible to choose a symmetric, smooth and periodic gauge \cite{Po2018, Po2018faithful, StiefelWhitney,Song}. We will generally always choose the gauge to be symmetric which means that it is either singular/discontinuous or not periodic. In general, this means that
\beq
 u_{\tau,n;\bG,a}(\bk+\bG_0) = \sum_m [U_{\tau,\bG_0}(\bk)]_{mn} u_{\tau,m;\bG + \bG_0,a}(\bk), \qquad U_{\tau,\bG_0}(\bk)^\dagger U_{\tau,\bG_0}(\bk) = 1
\eeq
where $U_{\tau,\bG}(\bk) = 1$ for any periodic gauge. This, in turn, implies
\beq
c^\dagger(\bk+\bG) = U_\bG(\bk) c^\dagger(\bk),
\qquad U_\bG(\bk) = \diag(U_{+,\bG}(\bk),U_{-,\bG}(\bk))_\tau
\eeq
Note that the momentum argument for $f^\dagger$ is unconstrained since we are using the continuum theory for monolayers of graphene and the normalization is chosen such that $\{f_{\mu,a}(\bq),f^\dagger_{\mu',a'}(\bq')\}=\delta_{\mu\mu'} \delta_{aa'} \delta_{\bq\bq'}$ (suppose the system size is finite), and $\inner{u_{\tau,n}(\bk)}{u_{\tau',n'}(\bk)}=\delta_{\tau \tau'}\delta_{nn'}$, which imply $\{ c_{\mu,n}(\bk),c^\dagger_{\mu',n'}(\bk')\}=\delta_{\mu\mu'}\delta_{nn'}\delta_{\bk\bk'}$ when $\bk,\bk'$ are confined in the MBZ. For the purpose of projecting the interaction into these two bands, it is convenient to introduce the form factor matrix 
\begin{align}
	[\Lambda_\bq(\bk)]_{\alpha,\beta} := \inner{ u_{\alpha}(\bk) } { u_{\beta}(\bk + \bq) } 
\end{align}
It follows from the definition that the form factor satisfies
\begin{align}
	\Lambda_\bq(\bk)^\dagger = \Lambda_{-\bq}(\bk + \bq), \qquad \Lambda_\bq(\bk + \bG) = U_\bG^*(\bk) \Lambda_\bq(\bk) U^T_\bG(\bk + \bq)
	\label{FFconstraint}
\end{align}

The interacting Hamiltonian is given by
\beq
\H_{\rm int} = \sum_\bk c^\dagger(\bk) h(\bk) c(\bk) - \frac{1}{2A} \sum_{\bq} V(\bq) :\rho_{\bq}  \rho_{-\bq}:, \qquad \rho_\bq = \sum_\bk c^\dagger(\bk) \Lambda_\bq(\bk) c(\bk + \bq)
\label{Hint}
\eeq
where $A$ is the total area of the system and $V(\bq)$ is the momentum space interaction potential, related to the real-space one by $V(\bq):= \int d^2 \br V(\br) e^{-i \bq \cdot \br}$ and $h(\bk)$ includes both the single-particle BM Hamiltonian as well as band renormalization effects due to remote bands not included in the projection. Depending on the number of gates, $V(\bq)$ takes the following form in the SI units: 
\begin{align}
	V(\bq)= \frac{e^2}{2 \epsilon\epsilon_0 q}
	\begin{cases}
	 (1-e^{-2q d_s}), &(\text{single-gate})\\
	 \tanh(q d_s), &(\text{dual-gate})
	\end{cases}
\end{align}
where the screening length $d_s$ is the distance from the graphene plane to the gate(s). Unless otherwise stated, we will use the double-gate-screened expression with $\epsilon = 9.5$ and $d_s = 40$nm.

\section{Hartree-Fock}

Here we detail our implementation of the Hartree-Fock method, in particular our ``subtraction'' scheme for avoiding a double-counting of the mean-field interaction, and our prescription for projecting onto a finite number of bands.  Modulo a soon-to-be-discussed correction, the Hamiltonian is
\begin{gather}
\label{HeffS}
\H_{\rm eff} = \sum_{\bk \in \rm BZ} c_\bk^\dagger h_{\textrm{BM}}(\bk) c_\bk - \frac{1}{2A} \sum_\bq V_\bq :\rho_\bq \rho_{-\bq}:, \\
\rho_\bq = \!\!\sum_{\bk \in \rm BZ}\!\! c_\bk^\dagger \Lambda_\bq(\bk) c_{\bk + \bq}, \quad [\Lambda_\bq(\bk)]_{\alpha,\beta} \! = \langle u_{\alpha,\bk} | u_{\beta,\bk + \bq} \rangle
\label{LambdaqkS}
\end{gather}
where $h_{\textrm{BM}}$ is the BM-Hamiltonian, with eigenstates labelled by $\alpha$. For the numerical calculations, we find it convenient to adopt a periodic gauge which means --using the notation of the previous appendix-- that $U_{\bG}(\bk)=\mathds{1}$ (note that in our analytical discussions we sometimes use a different gauge, namely a symmetric one).

Given a Slater determinant with correlation matrix $P_{\alpha \beta}(\bk) = \langle c_{\alpha, \bk}^\dagger c_{\beta, \bk} \rangle$, the corresponding Coulomb contribution to the Hartree-Fock Hamiltonian is
\begin{align}
H^{\textrm{C}}_{\textrm{MF}}[P](\bk) =  \frac{1}{A} \! \sum_{\bG} V_\bG  \Lambda_\bG(\bk) \sum_{\bk' \in \rm BZ} \tr \left( P(\bk') \Lambda^*_\bG(\bk')\right) - \frac{1}{A} \sum_{\bq} V_{\bq} \Lambda_{\bq}(\bk) P^T(\bk + \bq)  \Lambda^\dagger_{\bq}(\bk)\, ,
\end{align}
so that the total energy of this state is given by
\begin{align}
E_{\textrm{MF}} &= \sum_{\bk} \mbox{tr} \left(P(\bk) (h_{\textrm{BM}}(\bk) + \frac{1}{2} H^{\textrm{C}}_{\textrm{MF}}[P](\bk) )^T \right)
\end{align}
However, as pointed out in Ref. \cite{Liu19}, if the parameters of $h_{\textrm{BM}}$ are obtained by a method such as DFT, or by comparison with experiment, then $h_{\textrm{BM}}$ will already contain, to some extent, the effect of the interactions, and  the above expression will double-count this contribution. Consider, for example, the case where the two layers are decoupled, so that $h_{\textrm{BM}}$ is two copies of graphene. If we take for $P(\bk)$ the ground state of graphene at neutrality, then the Fock contribution to $H^{\textrm{C}}_{\textrm{MF}}[P]$ will lead to a logarithmically divergent renormalization of the Dirac velocity. However, the tight-binding parameters $h_{\textrm{BM}}$ are already chosen to replicate the measured Dirac velocity, so this renormalization will be unphysical.

To remedy this, it was suggested that the BM Hamiltonian should be replaced by $h(\bk) = h_{\textrm{BM}} - \frac{1}{2} H^{\textrm{C}}_{\textrm{MF}}[P^0](\bk)$, where $P^0$ is a ``reference'' density matrix such that  $h_{\textrm{BM}}$ is the full effective Hamiltonian when $P = P^0$.
The choice of $P^0$ then in principle depends on the method used to derive $h_{\textrm{BM}}$.
As in Ref.~\cite{Xie2018}, we choose $P^0$ to be the density matrix of two \emph{decoupled} graphene layers at neutrality.
While it may be tempting to choose $P^0$ to be the density matrix of $h_{\textrm{BM}}$ at neutrality, in most ab-initio methods \cite{Jeil2014} the parameters in $h_{\textrm{BM}}$ are obtained without any reference to the twist angle $\theta$, so it wouldn't make sense for $P^0$ to then depend on $\theta$.

Having chosen $P^0$, we must truncate to a finite number of bands for computational purposes.
We truncate based on projection into the eigenbasis $\alpha$ of $h_{\textrm{BM}}$, choosing $4N_- \leq \alpha \leq 4N_+$ of the bands closest to the flat bands ($N_-$ and $N_+$ denote the number of bands per spin and valley).
We assume that below / above $4N_- / 4N_+$ the density matrix is empty /  full, e.g.  $P_{\alpha \beta}(\bk) = \delta_{\alpha \beta}$ for $\alpha, \beta < 4N_-$ and $P_{\alpha \beta}(\bk) = 0$ for $\alpha, \beta > 4N_+$, while for $4N_- \leq \alpha, \beta \leq 4N_+$, $P_{\alpha \beta}(\bk)$ is determined by HF. 
In principle, this implies the HF Hamiltonian includes a contribution from all the filled bands $\alpha < N_-$.
However, there is also the corresponding subtraction of the reference density matrix $P^0_{\alpha \beta}$.
Because of the small inter-layer tunneling $w \sim 100$ meV, the contributions from $P_{\alpha\alpha}$ and $P^0_{\alpha\alpha}$ cancel out for $\alpha$ corresponding to bands far away from the charge neutrality point. 

With the subtraction of the reference density matrix $P^0$ taken into account, the Hartree-Fock mean field Hamiltonian is given by
\beq
\mathcal{H}_{\rm MF}[P] = \sum_{\bk}  c^\dagger_{\bk}\left(h(\bk) +H^C_{\rm MF}[P](\bk)  \right) c_{\bk} - \frac{1}{2}\sum_{\bk} \text{tr}\left(H^C_{\rm MF}[P](\bk)P(\bk)^T \right)
\eeq
The zero-temperature Hartree-Fock self-consistency condition states that the correlation matrix of the ground-state Slater determinant of $\mathcal{H}_{MF}[P]$ should be given by $P(\bk)$. To numerically solve the self-consistency equation we used both the `ODA' and `EDIIS' algorithms, both of which are developed and explained in detail in Refs.~\cite{RCA,ODA}.

\section{Flat band projected Hamiltonian}
Motivated by the numerical observation that mixing between the two flat bands and the remaining bands is relatively small for symmetry-broken phases at CN, we will only keep these two bands in the following discussion. The effect of the other bands will be included only through renormalization effects of the single-particle Hamiltonian $h(\bk)$ following the scheme of Ref.~\cite{Liu19}. 

\subsection{Sublattice-polarized basis}
In the chiral limit $w_0 = 0$, the sublattice operator $\sigma_z$ anticommutes with the BM Hamiltonian leaving the space of states spanning the flat bands invariant. Thus, we can choose the flat band states to be eigenstates of the sublattice operator $\sigma_z$. This basis is distinct from the band basis where the chiral symmetry operator is off-diagonal since it maps positive energy states to negative energy states. We note that the sublattice basis remains well-defined in the flat band limit for which the band basis is not well-defined.

\begin{figure}[h]
\includegraphics[width=0.7 \textwidth]{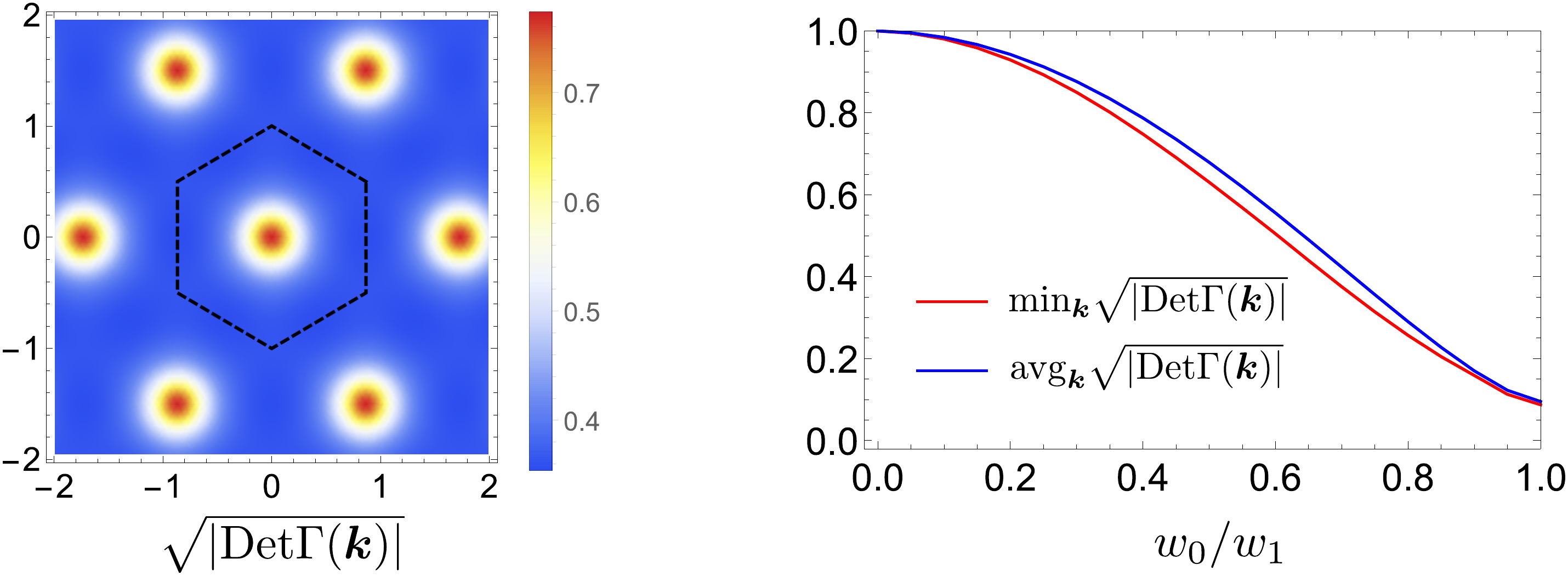}
\caption{Left panel shows the value of sublattice polarization computed by projecting the sublattice operator $\sigma_z$ on the flat band as a function of $\bk$ for the realistic parameters (\ref{parameters}). Right panel shows the minimum and average value of the sublattice polarization as a function of the ratio $w_0/w_1$. We can see that it never goes to zero indicating that the sublattice-polarized wavefunctions for the realistic model are adiabatically connected to those of the chiral model.}
\label{fGamma}
\end{figure}

Away from the chiral limit, we can still define the sublattice basis by diagonalizing the operator $\Gamma_{nm}(\bk) = \langle u_n(\bk)|\sigma_z|u_m(\bk) \rangle$. This yields a well-defined basis as long as the eigenvalues of $\Gamma(\bk)$ (which have equal magnitude and opposite sign due to $C_2 \T$) are non-zero, indicating a finite sublattice polarization. The sublattice polarization given by $\sqrt{|\det \Gamma(\bk)|}$ is plotted in the left panel of Fig.~\ref{fGamma} for the realistic model parameters (\ref{parameters}) and we can see it never goes to zero. This can also be seen in the right panel where the minimum and average value of sublattice polarization over the Moir\'e Brillouin zone is plotted as a function of $w_0/w_1$. We can clearly see from the plot that this value never geso to zero showing that the sublattice-polarized wavefunctions, those which diagonalize $\sigma_z$, for the realistic model are adiabatically connected to those of the chiral model. As in the main text, we will use the same Pauli matrices $\boldsymbol \sigma$ both sublattice index and the band index for sublattice-polarized wave-functions. It should be noted, however, that for $w_0 \neq 0$, the wavefunctions labelled by $\boldsymbol \sigma$ are only partially polarized on one of the sublattices i.e. they have amplitude on both sublattices. We note that for the chiral model at the magic angle, the sublattice-polarized wavefunctions have an explicit form in terms of theta functions given in Ref.~\cite{Tarnopolsky}.

To obtain the implementation of the different symmetries we start by noting that the eigenstates for a given spin can be labelled by their eigenvalues  under $\tau_z$ (valley index) and $\sigma_z$ (sublattice index). The phases of the four different wavefunctions can be chosen arbitrarily. Such choices will affect the form of the remaining symmetries in this basis. Once the phase of the wavefunction in valley K, sublattice A is fixed, we can use two of the three symmetries $C_2$, $\T$ and $\P$ (or some combinations of them) to fix the phase for the other three wavefunctions. Since we will be mostly using time-reversal and particle-hole symmetries, we will choose to fix these as
\beq
\T = \tau_x \K, \qquad \P = \tau_z \sigma_y \K
\eeq
which are chosen such that $\T$ flips valley but not sublattice ($\{ \T, \tau_z\} = 0$, $[\T, \sigma_z] = 0$) and $\P$ flips sublattice but not valley ($\{ P, \sigma_z\} = 0$, $[\P, \tau_z] = 0$). This leads to simple forms for $\P \T$ and $R$ symmetries
\beq
\P \T = \tau_y \sigma_y, \qquad R = \tau_y \sigma_x
\eeq

Once these operators are fixed, we are not free to choose the form of $C_2$, e.g. $\sigma_x \tau_x$. To see this, consider the modified two-fold rotation symmetry $C'_2 = i \tau_z C_2 R$ which is diagonal in sublattice, $[C'_2,\sigma_z] = 0$, and valley, $[C'_2, \tau_z] = 0$, and commutes with $R$ and $\T$. As a result, $C'_2 = e^{i \theta(\bk)}$ for some angle $\theta(\bk)$ satisfying $\theta(-\bk) = -\theta(\bk)$. Thus, $\theta(\bk) = 0, \pi$ at any TRIM ($\Gamma$, $M$, $M'$ and $M''$). We now note that sublattice-polarized badns has Chern number $\pm 1$ which implies that the sum of $\theta(\bk)$ over all TRIMs should be an odd multiple of $\pi$ \cite{Bernevig2012}. As a result, $\theta(\bk)$ cannot be constant over the Brillouin zone and has to have non-trivial $\bk$-dependence. The representation of $C_2$ in the same basis can be easily obtained as
\beq
C_2 = \sigma_x \tau_x e^{i \theta(\bk)}
\eeq
The symmetry representations in the sublattice-polarized basis in the chosen gauge are summarized in Table \ref{Symmetries}.

\subsection{Properties of the form factors}
Let us start with decomposing the form factor into parts which commutes/anticommute with the chiral symmetry $\sigma_z$
\beq
\Lambda_\bq(\bk) = \Lambda^S_\bq(\bk) + \Lambda^A_\bq(\bk), \qquad \Lambda^{S/A}_\bq(\bk) = \frac{1}{2}(\Lambda_\bq(\bk) \pm \sigma_z  \Lambda_\bq(\bk) \sigma_z) 
\label{Lambda}
\eeq
In the chiral limit, the space of flat bands is invariant under chiral symmetry leading to vanishing $ \Lambda^A_\bq(\bk)$.

Let us now see what is the most general form of $\Lambda^{S/A}_\bq(\bk)$ deduced from the other symmetries. $\U(2)_K \times \U(2)_{K'}$ imply that $\Lambda_\bq(\bk)$ is diagonal in valley and independent of spin, limiting it to the terms $\tau_{0,z} \sigma_{0,x,y,z}$. $\P \T$ symmetry further restricts these to $\tau_0 \sigma_{0,y}$ and $\tau_z \sigma_{x,z}$ whereas $C_2 \T$ enforces the term multiplying $\tau_z \sigma_z$ to be purely imaginary and the remaining terms to be purely real, leading to
\beq
 \Lambda^S_\bq(\bk) = F^S_\bq(\bk) e^{i \Phi^S_\bq(\bk) \sigma_z \tau_z}, \quad \Lambda^A_\bq(\bk) = \sigma_x \tau_z F^A_\bq(\bk) e^{i  \Phi^A_\bq(\bk) \sigma_z \tau_z}, 
\label{Fpm}
\eeq
In addition, $\T$ implies 
\beq
F^S_{-\bq}(-\bk) = F^S_{\bq}(\bk), \quad \Phi^S_{-\bq}(-\bk) = \Phi^S_{\bq}(\bk), \quad  F^A_{-\bq}(-\bk) = - F^A_{\bq}(\bk), \quad  \Phi^A_{-\bq}(-\bk) = \Phi^A_{\bq}(\bk)
\label{LambdaSym}
\eeq

\subsection{Hierarchy of scales}
\subsubsection{Particle-hole symmetric Hamiltonian}
The full Hamiltonian is given by (\ref{Hint}) where $h(\bk)$ is the renormalized single-particle dispersion. That is, the dispersion when the  flat bands are completely empty, i.e. $\nu = -4$. This can be written in terms of the dispersion at charge neutrality as
\beq
h(\bk) = h_{\nu=0}(\bk) - \frac{1}{A} \sum_\bG V_\bG \Lambda_\bG(\bk) \sum_{\bk'} \tr P_0(\bk') \Lambda^*_\bG(\bk') + \frac{1}{A} \sum_\bq V_\bq \Lambda_\bq(\bk) P_0^T(\bk + \bq) \Lambda^\dagger_\bq(\bk)
\label{Background}
\eeq
This expression assumed a periodic gauge such that $\Lambda_\bG(\bk)^\dagger = \Lambda_{-\bG}(\bk)$. Here, we have assumed there is always a solution to the HF equations at charge neutrality $\nu = 0$ which does not break any symmetry, with the self-consistent HF dispersion denoted by $h_{\nu=0}(\bk)$ and the projection onto the filled bands denoted by $P_0$. It can be numerically checked that such solution reproduces to a very good approximation the projection of the BM Hamiltonian on the two flat bands as suggested in Ref.~\cite{Liu19}. $\U_V(1)$, $C_2 \T$ and $\P \T$ imply that $h_{\nu = 0}$ has the form
\beq
h_{\nu=0}(\bk) =  a(\bk) \tau_z + f(\bk) \sigma_x e^{i \phi_0(\bk) \sigma_z \tau_z}
\eeq
which leads to
\beq
P_0(\bk) = \frac{1}{2}[1 + Q_0(\bk)], \qquad Q_0(\bk) = \sigma_x e^{i \phi_0(\bk) \sigma_z \tau_z}
\eeq
Substituting in (\ref{Background}), we get
\beq
h(\bk) = h_{\rm BM}(\bk) - \frac{1}{2A} \sum_\bG V_\bG \Lambda_\bG(\bk) \sum_{\bk'} \tr \Lambda^*_\bG(\bk') + \frac{1}{2A} \sum_\bq V_\bq \Lambda_\bq(\bk)  \Lambda^\dagger_\bq(\bk) + \frac{1}{2A} \sum_\bq V_\bq \Lambda_\bq(\bk) Q_0(\bk + \bq) \Lambda^\dagger_\bq(\bk)
\label{hCN}
\eeq
where we used the fact that $\tr Q_0(\bk) \Lambda^*_G(\bk)$ vanishes due to the form of $\Lambda_\bq(\bk)$ given in (\ref{Lambda}) and (\ref{Fpm}). Substituting in the interaction Hamiltonian, we get
\begin{gather}
\H_{\rm int} = \sum_\bk c_\bk^\dagger \tilde h(\bk) c_\bk + \frac{1}{2A} \sum_{\bq} V_\bq \delta \rho_\bq \delta \rho_{-\bq}^\dagger + \text{const.}\\
\tilde h(\bk) =  h_{\nu=0}(\bk)  + \frac{1}{2A} \sum_\bq V_\bq \Lambda_\bq(\bk) Q_0(\bk + \bq) \Lambda^\dagger_\bq(\bk), \qquad \delta \rho_\bq = \rho_\bq -\bar \rho_\bq, \qquad \bar \rho_\bq = \frac{1}{2} \sum_{\bG,\bk} \delta_{\bG,\bq} \tr \Lambda_\bG(\bk)
\label{HintCN}
\end{gather}
To reach this expression, we note that the normal-ordered interaction in (\ref{Hint}) differs from the density-density interaction in (\ref{HintCN}) by a bilinear term which cancels exactly against the third term in (\ref{hCN}). In addition, we separated the terms in the sum over $\bq$ corresponding to a reciprocal lattice vector $\bG$ and combined it with the second term in (\ref{hCN}). The advantage of this form of the Hamiltonian is that both terms are manifestly particle-hole symmetric.

\subsubsection{Estimation of energy scales}
Let us now write
\beq
\delta \rho_\bq = \delta \rho^S_\bq + \delta \rho^A_\bq, \qquad \delta \rho^{S/A}_\bq = \rho^{S/A}_\bq - \bar \rho^{S/A}_\bq, \qquad \rho^{S/A}_\bq =  \sum_\bk c_\bk^\dagger \Lambda^{S/A}_\bq(\bk) c_\bk, \qquad \bar \rho^{S/A}_\bq =  \frac{1}{2} \sum_{\bk,\bG} \delta_{\bG,\bq} \tr \Lambda^{S/A}_\bG(\bk)
\eeq
which induces a splitting of the interaction term into a symmetric intrasublattice part (under the unitary symmetry $R$) and a symmetry breaking intersublattice part given by
\beq
\H_S = \frac{1}{2A} \sum_{\bq} V_\bq \delta \rho^S_\bq \delta \rho^S_{-\bq}, \qquad \H_A = \frac{1}{2A} \sum_{\bq} V_\bq (\delta \rho^A_\bq \delta \rho^S_{-\bq}+ \delta \rho^S_\bq \delta \rho^A_{-\bq} + \delta \rho^A_\bq \delta \rho^A_{-\bq})
\eeq
We note here that although $\delta \rho^A \delta \rho^A$ does not by itself break the $R$ symmetry, it is only non-vanishing if this symmetry is broken since otherwise the antisymmetric form factor $\Lambda_\bq^A(\bk)$ vanishes.

\begin{figure}
\includegraphics[width=0.75 \textwidth]{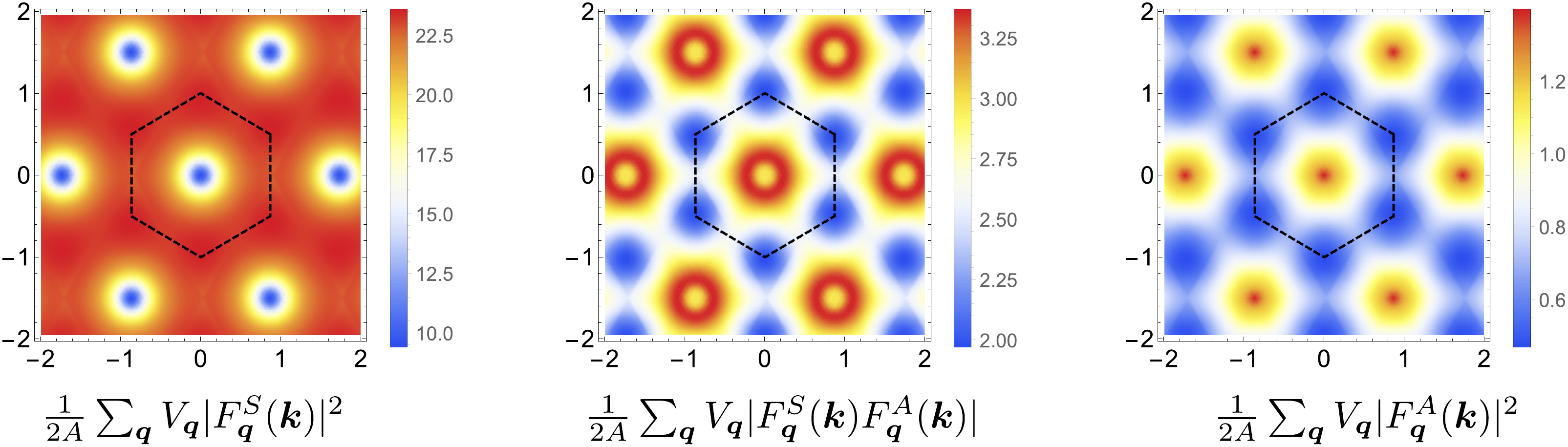}
\caption{The energy scale for the symmetry and symmetry breaking terms in the Hamiltonian obtained by averaging the squares of the form factors of the form factor $|F_\bq^S(\bk)|^2$, $F_\bq^S(\bk) F_\bq^A(\bk)$, $|F_\bq^A(\bk)|^2$ over $\bq$ weighted by the interaction as a function of $\bk$.}
\label{fVF}
\end{figure}

One crucial observation is that the magnitude of the symmetric form factor $F_\bq^S(\bk)$ is significantly larger than the magnitude of the antisymmetric form factor $F_\bq^S(\bk)$. This is shown in Fig.~\ref{fVF} where we plot the $\bq$ average of $|F_\bq^S(\bk)|^2$, $F_\bq^S(\bk) F_\bq^A(\bk)$, $|F_\bq^A(\bk)|^2$ weighted by the interaction. We see that the average values the three terms are about 17, 2 and 0.5 meV, respectively for the parameters in (\ref{parameters}). Although the absolute value of these terms depends on the interaction strength (the chosen value of $\epsilon$), the ratio between them is mostly sensitive to the properties of the wavefunctions close to the magic angle. It is worth noting that these ratios also depends slightly on the screening length as we will discuss later.

The $\bq$ averaged form factors can be used to define a characteristic scale for $\H_S$ and $\H_A$ by estimating the maximum absolute value of the exchange (Fock energy) as \cite{Liu19}
\beq
%\frac{1}{2N} \langle \Psi| \H_S | \Psi \rangle \leq 
U_S = \frac{1}{2A N} \sum_{\bk,\bq} V_\bq |F^S_\bq(\bk)|^2, \qquad %\frac{1}{2N} \langle \Psi| \H_A | \Psi \rangle \leq 
U_A = \frac{1}{2A N} \sum_{\bk,\bq} V_\bq \{ 2|F^S_\bq(\bk) F^A_\bq(\bk)| + |F^A_\bq(\bk)|^2\}
\label{Epm}
\eeq
which are plotted in Fig.~\ref{fVF} for the parameters in (\ref{parameters}). By averaging the plotted functions over $\bk$, we get the values of the bounds $U_S$ and $U_A$ to be 18, and 4.5 meV, respectively, for the choice of parameters (\ref{parameters}).

\begin{figure}
\includegraphics[width=0.5 \textwidth]{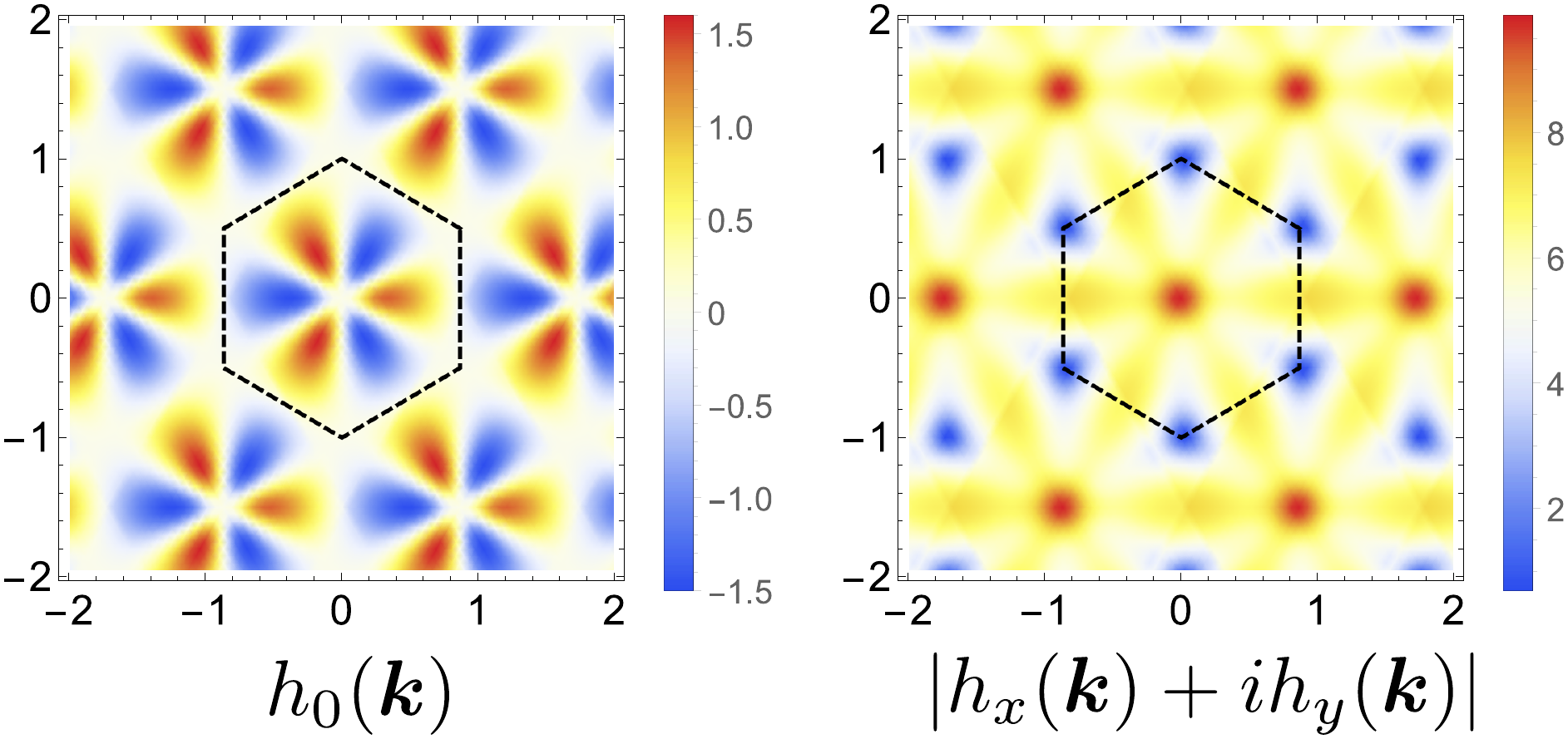}
\caption{The parameters of the renormalized dispersion defined in (\ref{h0xy}). We notice that the chiral symmetric part $h_{x,y}(\bk)$ is much larger than the symmetry breaking part $h_0(\bk)$.}
\label{fh0xy}
\end{figure}

Similarly, we can split $\tilde h(\bk)$ into a symmetric intersublattice part $\propto \sigma_x, \sigma_y \tau_z$ and a symmetry breaking intrasublattice part $\propto \tau_z$ as
\beq
\tilde h(\bk) = h_0(\bk) \tau_z + h_x(\bk) \sigma_x + h_y(\bk) \sigma_y \tau_z
\eeq
with
\begin{gather}
h_0(\bk) = \frac{1}{8} \tr \tilde h(\bk) \tau_z = a(\bk) + \frac{1}{A} \sum_\bq V_\bq F^S_\bq(\bk) F^A_\bq(\bk) \cos [\phi_0(\bk) - \Phi^S_\bq(\bk) - \Phi^A_\bq(\bk)] \nonumber \\
h_x(\bk) = \frac{1}{8} \tr \tilde h(\bk) \sigma_x = f(\bk) \cos \phi_0(\bk) + \frac{1}{A} \sum_\bq V_\bq \{ [F^S_\bq(\bk)]^2 \cos [\phi_0(\bk) - 2 \Phi^S_\bq(\bk)] + [F^A_\bq(\bk)]^2 \cos [\phi_0(\bk) - 2 \Phi^A_\bq(\bk)] \} \nonumber \\
h_y(\bk) = \frac{1}{8} \tr \tilde h(\bk) \sigma_y \tau_z = f(\bk) \sin \phi_0(\bk) + \frac{1}{A} \sum_\bq V_\bq \{ [F^S_\bq(\bk)]^2 \sin [\phi_0(\bk) - 2 \Phi^S_\bq(\bk)] - [F^A_\bq(\bk)]^2 \sin [\phi_0(\bk) - 2 \Phi^A_\bq(\bk)] \}
\label{h0xy}
\end{gather}
We note that, the interaction-induced renormalization of the symmetric piece of the dispersion $h_{x,y}$ contains the symmetric form factor $F^S_\bq(\bk)$ whereas the asymmetric piece $h_0$ contains at least one factor of the asymmetric form factor $F^A_\bq(\bk)$. Thus our previous discussion (cf.~Fig.~\ref{fVF}) implies that $h_{x,y}$ is, on average, significantly larger than $h_0$. This is verified by plotting the values of $h_0(\bk)$ and $|h_x(\bk) + i h_y(\bk)|$ ($h_{x,y}(\bk)$ are not separately gauge invariant) as a function of $\bk$ as shown in Fig.~\ref{fh0xy}. By averaging  $|h_0(\bk)|$ and $|h_x(\bk) + i h_y(\bk)|$, we can obtain an estimate for the energy scales associated with the symmetric and symmetric breaking piece of the dispersion
\beq
t_S = \frac{1}{N} \sum_\bk |h_x(\bk) + i h_y(\bk)|, \qquad t_ A= \frac{1}{N} \sum_\bk |h_0(\bk)|
\eeq
leading to $t_S \approx$ 5 meV and $t_A \approx$ 0.5 meV.

The previous discussion implies that the intrasublattice part of the interaction $\H_S \sim$ 20 meV is the largest scale in the problem. It is followed by the intersublattice part of the dispersion $h_{x,y}(\bk)$ and the intersublattice part of the interaction $\H_A$ which are of the same order $\sim 5$ meV which is about a factor of 4 smaller than $\H_S$. The non-symmetric part of the dispersion is much smaller $\sim 0.5$ meV. 

\subsubsection{Dependence on screening}

\begin{figure}
    \centering
    \includegraphics[width=0.9\textwidth]{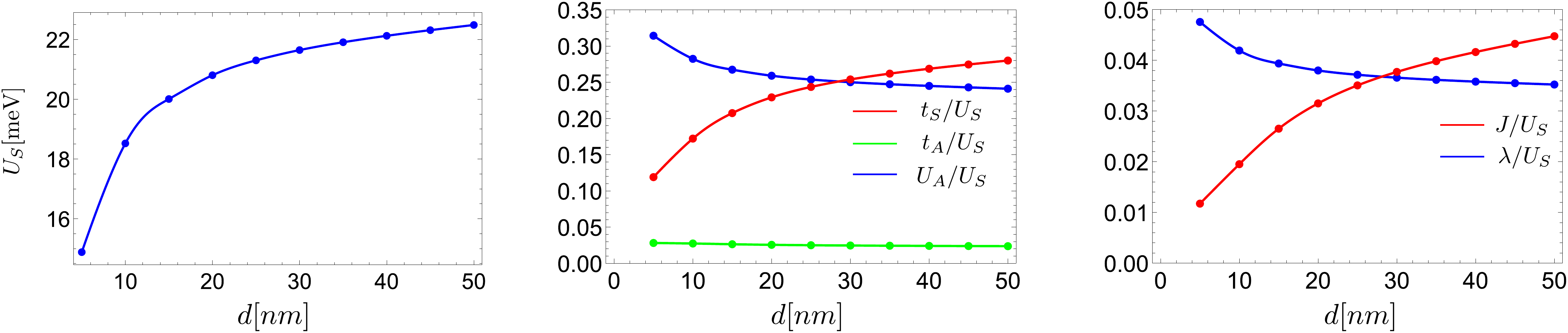}
    \caption{Dependence of the different energy scales on the screening length $d$. The left panel shows the overall interaction scale in meV. The middle panel shows the strength of the symmetric dispersion $t_S$, symmetry-breaking interaction $U_A$, and symmetry-breaking dispersion $t_A$ relative to the symmetric interaction scale $U_S$. The right panel shows the strength of the antiferromagnetic coupling $J$ and the energy splitting of the ground state manifold due to the symmetry breaking interaciton $\lambda$.}
    \label{Screening}
\end{figure}

Although the relative strength of the different parameters $U_{S/A}$, $t_{S/A}$ is insensitive to the overall strength of the interaction determined by the dielectric constant $\epsilon$, it can be sensitive to the form of the interaction which is controlled by the screening length $d$ which was shown to be tunable in recent experiments \cite{EfetovScreening, YoungScreening} by tuning the distance to the metallic gate. Apart from the dependence of the overall interaction scale on the screening length $d$ (left panel in Fig.~\ref{Screening}), we can also see that the magnitude of the dispersion $t_S$ and symmetry breaking interaction $U_A$ relative to the symmetric interaction $U_S$ depend weakly on the screening length. In particular, although both scales are always a factor of 3-5 smaller than the interaction, their relative strength depends on the screening, with the dispersion favored by larger screening length and the symmetry breaking interaction favored by smaller screening length.

\subsubsection{Hierarchy of symmetries}
The largest scale in the problem $\H_S$ is associated with an enlarged symmetry which can be seen by noting its invariance under any unitary transformation
\beq
\label{U}
c \mapsto U c, \qquad [U, \sigma_z \tau_z] = 0
\eeq
which yields the symmetry group U(4)$\times$U(4) denoting independent rotations between states within a fixed Chern number sector. 

The discussion of symmetry is simplified if we introduce the a new basis $\tbtau$ and $\tbsigma$ such that $\frac{1}{2}(1 \pm \ttau_z)$ projects onto the space of bands with Chern number $\pm 1$ and $\tbsigma$ labels the states with the same spin within this subspace. This means the subspace of bands with $\ttau_z = 1$ consists of states polarized to sublattice A in the $+$ valley and sublattice B in the $-$ valley and vice versa for $\ttau_z = -1$ with $\tbsigma$ specifying the state within this 2-band subspace. More specifically, we define
\beq
\tbtau = (\sigma_x, \sigma_y \tau_z, \sigma_z \tau_z), \qquad \tbsigma = (\sigma_x \tau_x, \sigma_x \tau_y, \tau_z)
\eeq
In this basis, the symmetry of $\H_S$ is given by
\beq
U = \left(\begin{array}{cc} U_1 & 0 \\ 0 & U_2 \end{array} \right)_\ttau
\label{U12}
\eeq

This symmetry is broken in two different ways. The intersublattice dispersion $h_{x,y}(\bk)$ breaks it down to $\U(4)$ by requiring $U$ to commute with $\sigma_x = \ttau_x$ which is equivalent to the condition $U_1 = U_2$ in (\ref{U12}) reducing the $\U(4) \times \U(4)$ symmetry of the interaction to $\U(4)$.

The other symmetry breaking effect comes from the intersublattice interaction $\H_A$ which reduces the symmetry of the interaction from $\U(4) \times \U(4)$ to $\U(4)$ by enforcing the extra condition that $U$ commutes with $\sigma_x \tau_z = \tsigma_z \ttau_x$ leading to $U_1 = \tsigma_z U_2 \tsigma_z$. This $\U(4)$ subgroup is different from the $\U(4)$ subgroup leaving $h_{x,y}(\bk)$ invariant. In the presence of both $h_{x,y}(\bk)$ and $\H_A$, the symmetry is obtained by intersecting the two $\U(4)$ subgroups leading to $U_1 = U_2 = \tsigma_z U_1 \tsigma_z$ which implies
\beq
U_1 = U_2 = \left(\begin{array}{cc} V_+ & 0 \\ 0 & V_- \end{array} \right)_\tsigma
\eeq
This corresponds to $\U(2)_K \times \U(2)_{K'}$ symmetry corresponding to independent $\U(2)$ rotation within each valley. We notice that the much smaller intrasublattice dispersion $h_0$ does not introduce any extra symmetry breaking.

\section{Symmetric strong coupling limit}
Motivated by the discussion of the previous section, we will now consider a strong coupling approach to the problem by assuming that the intrasublattice interaction scale is much larger than both the intersublattice interaction $\H_A$ and the intersublattice single-particle Hamiltonian $h_{x,y}$, i.e. $U_S \gg U_A, t_S$. Furthermore, we will neglect the intrasublattice dispersion $h_0$ since it is smaller than all other terms and does not break any symmetry that is unbroken at a higher energy scale.

\subsection{Spinless model}
For simplicity, let us first consider the simpler problem of spinless electrons at half-filling with the $\U(4) \times \U(4)$ symmetry replaced by $\U(2) \times \U(2)$. We start by noting that since $\delta \rho^S_{-\bq} = [\delta \rho^S_\bq]^\dagger$, the Hamiltonian $\H_S$ is a non-negative definite operator. As a result, any many-body state annihilated by $\H_S$ is a ground state \cite{Repellin19, Alavirad19}. We further note that for any Slater determinant state $|\Psi_Q\rangle$ described by the density matrix $P(\bk) = \frac{1}{2}(1 + Q)$ where $Q$ commutes with $\sigma_z \tau_z$, the action of the Hamiltonian is given by
\beq
\H_S |\Psi_Q\rangle = 2N E_C |\Psi_Q\rangle, \qquad E_C = \frac{C_T^2}{A N} \sum_\bG V_\bG |\sum_\bk F^S_\bG(\bk) \sin \Phi^S_\bG(\bk)|^2 
\eeq
where $C_T$ is the total Chern number of the state. In a periodic gauge, (\ref{FFconstraint}) implies $\Lambda_\bG(\bk) = \Lambda^\dagger_{-\bG}(\bk)$ which, together with (\ref{LambdaSym}), implies that $\Phi^S_\bG(\bk) = -\Phi^S_{-\bG}(\bk)$. As a result, the summation over $\bk$ in $E_C$ vanishes identically, leading to $\H_S |\Psi_Q\rangle = 0$. Thus, we see that the states $|\Psi_Q \rangle$ with $Q$ commuting with $\sigma_z \tau_z$ are exact ground states of $\H_S$. 

These states can be understood using the schematic illustration of Fig.~3 in the main text. The figure contains four bands labelled by valley and sublattice indices and since the symmetric form factor $\Lambda^S_\bq(\bk)$ is diagonal in both, the interaction $\H_S$ is minimized by minimizing density fluctuations within each band which is achieved by completely filling two of the four bands. There are several possible ways to do this which can be grouped into two categories: (i) filling two bands with the same Chern number leading to a $C_T = \pm 2$ QH states which is invariant under $\U(2) \times \U(2)$, or (ii) filling two states with opposite Chern number leading to a manifold of $C_T = 0$ states generated by acting with $\U(2) \times \U(2)$ on the VP state $Q = \tau_z$.

\subsubsection{Effect of $h_{x,y}$}
Next, we consider the effect of the dispersion $h_{x,y}(\bk)$ on this manifold of states. This breaks the symmetry down to $\U(2)$ by coupling pairs of bands with opposite Chern number through tunneling with amplitude $h_x(\bk) + i h_y(\bk)$. This can be seen by noting that the symmetric dispersion has the form $\ttau_x h_x(\bk) + \ttau_y h_y(\bk)$, thus, in the $\ttau,\tsigma$ basis, it connects bands with opposite $\ttau_z = \pm 1$ with the same value of $\tsigma_z$. 

We first note that, among the manifold of ground states of $\H_S$, those for which $Q$ commutes with $\sigma_x$ are annihilated by $h_{x,y}$. These correspond to states in which each pair of bands connected by the tunelling in Fig.~4 are both either filled or empty, forming the manifold $\U(2) / \U(1) \times \U(1) \simeq S^2$ spanned by the VP state $Q = \tau_z$ and $\T$-IVC. To understand the action of $h_{x,y}$, let us first consider the case in which $Q$ anticommutes with $ h_{x,y}$ corresponding to states for which only one out of each pair of coupled bands in Fig.~4 is filled. The action of $ h_{x,y}$ on such states creates an electron-hole pair by moving an electron at momentum $\bk$ from a filled band to the same momentum at the corresponding empty band with opposite Chern number.

The energy associated with such process can be calculated within second order perturbation theory where an electron-hole pair is created then annihilated. Since this can be done for each pair of bands independently, we can limit ourselves to only one pair of bands with opposite Chern number coupled by $h_{x,y}$ which can be labelled with $C = \sigma_z \tau_z = \ttau_z = \pm 1$. We denote a state with an electron-hole pair with momentum $\bk$ by
\beq
 |\Psi_{\bk,\rm eh} \rangle = c_{\bk,-}^\dagger c_{\bk,+} |\Psi_+ \rangle, \qquad |\Psi_+ \rangle = \prod_{\bk} c_{\bk,+}^\dagger |0\rangle
\eeq
where $|\Psi_+ \rangle$ is the state where the $C = +1$ band is filled and the $C=-1$ band is empty. The energy correction (per particle) due to the coupling $h_x(\bk) + i h_y(\bk)$ between the two bands is then given by
\beq
\Delta E = -J = -\frac{1}{N}\sum_{\bk,\bk'} [h_x(\bk) + i h_y(\bk)] [\H_{\rm eh}^{-1}]_{\bk,\bk'} [h_x(\bk') - i h_y(\bk')], \qquad [\H_{\rm eh}]_{\bk,\bk'} = \langle \Psi_{\bk,\rm eh}| \H_S | \Psi_{\bk', \rm eh} \rangle
\eeq
We note that operator $\H_S$ conserves the number of electron-hole pairs and their momentum so it acts within the space of states $|\Psi_{\bk, \rm eh} \rangle$. As a result, $\H_{\rm eh}$ is given by 
\beq
[\H_{\rm eh}]_{\bk,\bk'} = \frac{1}{A} \sum_\bq V_\bq |F^S_\bq(\bk)|^2 \{\delta_{\bk,\bk'} - \delta_{\bk',[\bk + \bq]} e^{2i \phi_\bq(\bk)} \}
\label{Heh}
\eeq
where $[\bq]$ denotes the part of $\bq$ within the first zone. This expression can be understood by noting that the action of $\delta \rho^S_\bq$ on an electron-hole pair either increases the momentum of the electron by $\bq$ or decreases the momentum of the hole by $\bq$. As a result, the action of $\delta \rho^S_{-\bq} \delta \rho^S_\bq$ either returns the electron-hole pair to its initial state or shifts both momenta by $\pm \bq$. The matrix $\H_{\rm eh}$ can be easily evaluated numerically leading to $J \approx 1.5$ meV. This agrees with the simple estimate which assumes that the typical value for the eigenvalues of $\H_{\rm eh}$ are of the same order as the interaction scale 15-20 meV, yielding $J \approx t_S^2/U_S$ 1-3 meV. The dependence of $J$ on the screening is shown in Fig.~\ref{Screening}. We can see that the antiferromagnetic coupling increases with increasing the screening length due to the increase in the dispersion renormalized dispersion $t_S$.

One important observation that is crucial for the previous argument is that the spectrum of the electron-hole pairs has a gap. This feature can be traced to the band topology as follows. We start by writing a general state with a single electron-hole pair
\beq
|\Psi_{\rm eh} \rangle = \frac{1}{\sqrt{N}}\sum_\bk a_\bk |\Psi_{\bk,\rm eh} \rangle, \qquad \sum_\bk |a_\bk|^2 = N
\label{Psi}
\eeq
We note that restricting this sum to the first Brillouin zone requires $a_\bk c_{\bk,-}^\dagger c_{\bk,+}$ to be periodic in $\bk$. Due to band topology, it is impossible to choose a smooth and periodic gauge. In the following discussion, we will prefer to choose a smooth gauge for which the operators $c_{\bk,\pm}$ are not periodic in $\bk$. Instead, their phase winds by $\pm 2\pi$ around the Brillouin zone. As a result, the phase of $a_\bk$ should wind by $4\pi$ around the Brillouin zone, i.e.  $a_\bk$ has at least two vortices.

Substituting in (\ref{Heh}), we get
\beq
\langle \Psi_{\rm eh}| \H_{\rm eh} | \Psi_{\rm eh} \rangle = \frac{1}{A N} \sum_{\bk,\bq} V_\bq [F^S_\bq(\bk)]^2 [|a_\bk|^2 - e^{2 i \Phi_\bq(\bk)} a^*_{\bk + \bq} a_\bk] %+ \frac{1}{2A} \sum_\bG V_\bG |F^S_\bG(\bk)|^2 \sin^2 \Phi_\bG(\bk)
\eeq
%The last term is a constant which vanishes in the thermodynamic limit and can be dropped. 
We now notice that to leading order in $\bq$, $\Phi^S_\bq(\bk)$ can be written as $\Phi^S_\bq(\bk) = -\bq \cdot \bA_\bk + O(\bq^3)$ where $\bA_\bk$ is the Berry connection
\beq
\bA = -i \langle u_\bk| \nabla| u_\bk \rangle, \qquad \frac{1}{2\pi} \int_{\rm BZ} \nabla_\bk \times \bA_\bk = 1
\label{Berry}
\eeq
 To make further analytical progress, we assume that the magnitude of the form factor $F^S_\bq(\bk)$ depends only on $|\bq|$ and decays relatively quickly in $\bq$ on the scale of the Brillouin zone leading to
\begin{gather}
\langle \Psi_{\rm eh}| \H_{\rm eh} | \Psi_{\rm eh} \rangle = \frac{c}{N} \sum_\bk  |(\nabla_\bk - 2 i \bA_\bk) a_\bk|^2 = c \int_{\rm BZ} \frac{d^2 \bk}{(2\pi)^2} |(\nabla_\bk - 2 i \bA_\bk) a_\bk|^2, \nonumber \\ c = \frac{1}{2A} \sum_{\bq \neq \bG} \bq^2 V_\bq |F^S_\bq|^2 = \frac{1}{2} \int \frac{d^2\bq}{(2\pi)^2} \bq^2 V_\bq |F^S_\bq|^2
\label{EGL}
\end{gather}
where $c$ is the constant of the same order as the interaction scale. We note that (\ref{EGL}) has the same form as the Ginzburg-Landau free energy of a superconductor in a magnetic field in momentum space which is in the vortex lattice phase with two vortices per unit cell due to the non-trivial winding of $a_\bk$. The free energy of such phase is always larger than zero which can be seen by employing similar arguments to those of Ref.~\cite{Bultinck19} as shown below.

We start with the observation that, in a smooth non-singular gauge, $\bA_\bk$ is finite everywhere in the Brillouin zone $|\bA_\bk| \leq |\bA_{\bk, \rm max}|$. Without loss of generality, we can write $a_\bk = \rho (k_x + i k_y)$ close to a vortex. If we choose a ball of radius $\epsilon \ll |\bA_{\bk, \rm max}|^{-1}$ around the vortex, then its energy is given by $\frac{c}{\pi} \rho^2 \epsilon^2 [1 + O(\epsilon^2 |\bA_{\bk, \rm max}|^2)]$ which is always finite since $\rho$ cannot be made arbitrarily small (this is a result of the normalization constraint  (\ref{Psi})). As a result, the energy expectation value in (\ref{EGL}) is always positive, i.e. the Hamiltonian $\H_{\rm eh}$ is gapped. 

In summary, $h_{x,y}$ favors states where only one from each pair of bands connected by $h_{x,y}$ is filled, allowing for virtual hopping which lowers their energy by an amount $J \sim t_S/U_S$. This is equivalent to the condition $\{Q,\sigma_x\}$ which selects a manifold of states consisting of two sectors: a $\U(2)$-invariant QH state with Chern number $\pm 2$ and a manifold of states with vanishing Chern number corresponding to $\U(2)/\U(1) \times \U(1) \simeq S^2$ generated by the VH and K-IVC states.

\subsubsection{Effect of $\H_A$}

The intersublattice interaction $\H_A$, on the other hand, picks a different submanifold of ground states corresponding to the states annihilated by the non-symmetric density operator $\delta \rho^A_\bq$. These are characterized by $Q$ which commutes with $\sigma_y$ forming the manifold $\U(2) /\U(1) \times \U(1) \simeq S^2$ generated by the VP and K-IVC. The energies of the remaining states for which $Q$ anticommutes with $\sigma_y$ (QH, VH, and $\T$-IVC)  are increased by an amount of the order 
\beq
\lambda = \frac{1}{2AN} \sum_{\bk,\bq} V_\bq |F^A_\bq(\bk)|^2 \approx 0.5 \text{meV}.
\eeq
The dependence of $\lambda$ on the screening is shown in Fig.~\ref{Screening}. We can see that it remains relative constant except for very small values of screening where it starts slightly increasing. 

Thus, in the presence of both $h_{x,y}$ and $\H_A$, the K-IVC, which benefits from both perturbations, has the lowest energy followed by the VP and QH/VH (the latter two are degenerate) whose competition is determined by the relative strength of the symmetry-breaking terms in the interaction $\lambda$ and the energy gain due to virtual tunneling $-J$. The $\T$-IVC state, which was not seen in the numerics, is disfavored by both and has the highest energy. 

\subsection{Spinful model}

\subsubsection{Charge neutrality}
Upon including the spin, we can study the manifold of ground states at CN in a similar fashion starting with the states which minimize $\H_S$ for which $Q$ commutes with $\sigma_z \tau_z$. These states are obtained by completely filling 4 of the 8 bands in Fig.~3. There are three sectors of such states with Chern numbers $\pm 4$, $\pm 2$ and $0$. The manifold of Chern number $\pm 4$ consists of a single state $Q = \sigma_z \tau_z$ invariant under $\U(4) \times \U(4)$ rotations which is the analog of the QH state in the spinless case. The manifold of states with Chern number $\pm 2$ is obained by filling three bands with one Chern number and a single band with the opposite Chern number. This manifold is 12-dimensional and is isomorphic to $[\U(4)/\U(3) \times \U(1)]^2$. The states within this manifold break several symmetries and are characterized by mixed orders where, for instance, one valley has a spin-polarized state whereas the other has a $C_2 \T$-breaking sublattice polarized state. They necessarily involve some non-trivial spin structure and do not have analogs in the spinless problem. Finally, the manifold of zero Chern number states is 16-dimensional and is isomorhphic to $[\U(4)/\U(2) \times \U(2)]^2$. This manifold includes all the zero Chern number states found in the spinless model (VH, VP, $\T$-IVC, and K-IVC) in their spin-singlet $Q \propto s_0$ or spin-triplet $Q \propto \bn \cdot \bs$ variants. It also includes the purely spin-polarized (SP) state $Q = \bn \cdot \bs$.

The dispersion $h_{x,y}$ selects for the states where only one band out of each pair of bands coupled by $h_{x,y}$ is filled which is equivalent to the requirement that $Q$ anticommutes with $\sigma_x$. This reduces the manifold of state with Chern number $\pm 2$ to a 6-dimensional manifold $\U(4) / \U(3) \times \U(1)$ and that of Chern number 0 states to an 8-dimensional manifold $\U(4)/ \U(2) \times \U(2)$. The latter includes the VH and K-IVC states (both spin-singlet and triplet versions) but does not include the SP, VP or $\T$-IVC states.

The non-symmetric interaction $\H_A$ selects a different submanifold of states with the requirement that $Q$ commutes with $\sigma_x \tau_z$. The possible states satisfying this requirement has Chern numbers $\pm 2$ or 0 with the former forming the 6-dimensional manifold $\U(4) / \U(3) \times \U(1)$ and the latter forming 8-dimensional manifold $\U(4)/ \U(2) \times \U(2)$. The zero Chern number states include the VP, SP, and K-IVC states.

Thus, the presence of both $h_{xy}$ and $\H_A$ selects the K-IVC state as the unique ground state up to the action of $\U(2)_K \times \U(2)_{K'}$ symmetry corresponding to independent spin rotation in each valley. The most general K-IVC state is given by
\beq
Q = \sigma_y \otimes \left(\begin{array}{cc} 0 & V \\ V^\dagger & 0 \end{array}\right)_\tau
\eeq
where $V$ is a 2 $\times$ 2 unitary matrix acting in the spin space. The gound state manifold is thus isomorphic to $\U(2)$. The action of $\U(2) \times \U(2)$ symmetry on the ground state is
\beq
V \mapsto U_+^\dagger V U_- 
\eeq
which generates the full $\U(2)$ group starting from any given state , e.g. $V = 1$. This means that any given state is invariant under a $\U(2)$ subgroup given by the condition $U_- = V^\dagger U_+ V$ which is consistent with the ground state manifold being $\U(2) \simeq \frac{\U(2) \times \U(2)}{\U(2)}$. This manifold which can be written as $\U(1) \times \SU(2)$ can be parametrized as $e^{i \phi} e^{i \frac{\theta}{2} \bn \cdot \bs}$, $0 \leq \phi < 2\pi$, $0 \leq \theta \leq \pi$. Here, $\phi$ and $\theta$ can be associated with the total and relative angles of the IVC orders in the spin up and down sectors.

In the presence of intervalley Hund's term given by
\beq
\H_J = \frac{J}{N} \bS_{+,\bq} \cdot \bS_{-,-\bq}, \qquad \bS_{\pm, \bq} = \sum_{\bk} c^\dagger_\bk \frac{1 \pm \tau_z}{2} \bs c_{\bk + \bq}
\eeq
the $\U(2)_K \times \U(2)_{K'}$ symmetry is broken down to $\U_C(1) \times \U_V(1) \times \SU(2)$. For the ground state $Q$ parametrized by $\phi$, $\theta$ and $\bn$, we note that the energy should remain independent on $\phi$ and $\bn$ since they still transform non-trivially under the symmetry. Thus, we can fix their value to $\phi = 0$ and $\bn = \hat z$. 
We now evaluate $E_J(\theta) = \langle \Psi_\theta | \H_J | \Psi_\theta \rangle$ which is nothing but the Hartree-Fock decoupling of $\H_J$. The Hartree term vanishes whereas the Fock term yields (up to a constant)
\beq
E_J(\theta) \propto - J \sum_a \tr e^{i \frac{\theta}{2} s_z} s^a e^{-i \frac{\theta}{2} s_z} s^a = -2J (1 + 2 \cos \theta)
\eeq
For ferromagnetic Hund $J<0$, the minimum is at $\theta = \pi$ while for antiferromagnetic Hund $J>0$, the minimum is at $\theta = 0$.

\subsubsection{Half-filling}
The ground state at half-filling $\nu = \pm 2$ can be understood similarly. In the following, we will focus on the case $\nu = -2$. The $\nu = 2$ case is very similar. At $\nu = -2$, $Q$ satisfies $\tr Q = - 4$ and can be generally written as
\beq
Q = -P_- + P_+ Q_1, \qquad P_\pm = \frac{1}{2}(1 \pm Q_2), \qquad Q_{1,2}^2 = 1, \quad \tr Q_{1,2} = 0, \quad [Q_1, Q_2] = 0
\eeq
This describes a state where the 4 bands associated with the projector $P_-$ are completely empty, whereas the 4 bands associated with the projector $P_+$ are half-filled.

The analysis of the ground state manifold is very similar to the CN case. We start by the states which minimize $\H_S$ which are specified by requiring $[Q_{1,2},\sigma_z \tau_z] = 0$. This is equivalent to filling 2 out of the 8 bands of Fig.~4. The manifold of ground state consists of two sectors depending on whether the two filled bands have the same Chern number. The first sector contains Chern number 2 states such as spin-polarized QH state $Q_2 = s_z$, $Q_1 = \sigma_z \tau_z$ which form the manifold $\U(4) / \U(2) \times \U(2)$. The second sector contains Chern number 0 states such as spin-polarized VH state $Q_2 = s_z$, $Q_1 = \sigma_z$ or spin and valley polarized states $Q_2 = s_z$, $Q_1 = \tau_z$ which form the manifold $[\U(4)/\U(3) \times \U(1)]^2$.

Including $h_{x,y}$ selects states for which at most one band from each pair connected by the tunneling $h_{x,y}$ is filled. This is equivalent to the condition $[Q_2, \sigma_x] = 0$, $\{Q_1, \sigma_x \}=0$. All states with Chern number 2 are included in this manifold. For the zero Chern number sector, this selects a submanifold of states isomorphic to $\U(4)/\U(3) \times \U(1)$. The non-symmetric part of the interaction $\H_A$ instead requires states related by the action of $\sigma_x \tau_z$ to be both filled or both empty. This is equivalent to the condition $[Q_{1,2}, \sigma_x \tau_z] = 0$. This condition rules out all states with non-vanishing Chern number and it selects a submanifold of zero Chern number states isomorphic to $\U(4) / \U(3) \times \U(1)$. 

The two constraints are only simultaneously satisfied by K-IVC states. The simplest such state at half-filling is the spin-polarized IVC state obtained by taking $Q_2 = s_z$ and $Q_1 = \sigma_y(\tau_+ e^{i \phi} + \tau_- e^{i \phi})$. The manifold of ground state is generated by acting with $\U(2)_K \times \U(2)_{K'}$ on this state yielding
\beq
Q = \left(\begin{array}{cc} \frac{1}{2}(1 + \bn_+ \cdot \bs) & \Xi \sigma_y \\ \Xi^\dagger \sigma_y & \frac{1}{2}(1 + \bn_- \cdot \bs) \end{array}\right)_\tau, \qquad \bn_\pm = \frac{1}{2} \tr U^\dagger_\pm s_z U_\pm \bs,\qquad \Xi = U_+^\dagger P_\downarrow U_-, \qquad P_{\uparrow/\downarrow} = \frac{1 \pm s_z}{2}
\eeq
where $U_+$, and $U_-$ are $2 \times 2$ matrices acting in spin space. This state is parametrized by $U_+$ and $U_-$. However, we note that the replacement $U_\pm \mapsto e^{i (\phi_\pm P_\downarrow + \phi P_\uparrow)} U_\pm$ for any phases $\phi$, $\phi_+$ and $\phi_-$ does not change the state $Q$. Thus, $Q$ parametrizes the manifold $\frac{\U(2) \times \U(2)}{\U(1) \times \U(1) \times \U(1)} \simeq \U(1) \times S^2 \times S^2$. Here, the $\U(1)$ denotes the IVC phase and the two $S^2$ factors denote the spin direction in each valley. Such states denotes a spin polarized IVC with the spin direction in each valley chosen independently.

The intervalley Hund's coupling will select the state where the spin is aligned in both valleys for $J<0$ which can be implemented by taking $U_+ = e^{i \phi} U_-$. For $J>0$, it will select the spin-valley-locked state where spins are anti-aligned in the two valleys. This is implemented by the choice $U_+ = s_x e^{i \phi} U_-$.

%\bibliography{refs}

\end{document}